\RequirePackage{etex}
\documentclass{aa}  
\usepackage{graphicx}
\usepackage{placeins}
\usepackage{lscape}
\usepackage{amsmath}
\usepackage{amssymb}
\usepackage{txfonts}
\usepackage{natbib}
\usepackage{hyperref}\usepackage{booktabs}
\hypersetup{colorlinks=true,linkcolor=blue,citecolor=blue,urlcolor=blue}
\usepackage{xcolor}
\usepackage{chngcntr}
\usepackage{rotating} 
\usepackage{booktabs}
\usepackage{graphicx}
\usepackage{orcidlink}
\makeatletter
\renewcommand{\@makefnmark}{\hbox{\@textsuperscript{\hyper@linkstart{link}{\@thefnmark}\color{blue}\@thefnmark\hyper@linkend}}}
\makeatother

\begin{document}

\title{XMM-Newton observations of the highest field polar AR UMa}
\subtitle{An extreme case of soft-blobby accretion}
\author{K.G. Pradeep\inst{1,2} 
         \and A.D. Schwope\inst{1}}
\institute{Leibniz-Institut für Astrophysik Potsdam (AIP), 
           An der Sternwarte 16, 14482 Potsdam, Germany\\
           \email{kpradeep@aip.de}
           \and Potsdam University, Institute for Physics and Astronomy, 
           Karl-Liebknecht-Straße 24/25, 14476 Potsdam, Germany}

\date{Accepted October 2026. Received September 2026}

   \abstract
 {The X-ray spectra of magnetic cataclysmic variables (MCVs) with extreme magnetic fields, the polars, often exhibit a prominent soft X-ray excess attributed to buried shocks.}
{Using the first-ever high-state \textit{XMM-Newton} observations of the highest field polar, AR~UMa (B$\approx$230~MG), we established the source's accretion scenario and derived the energy budget at the magnetic extreme of known MCVs.} 
{Using X-ray, ultraviolet, and optical spectral and timing analysis, we disentangled the radiation components in the system and located their origin in the binary.}
{AR UMa showed one active (``southern'') pole during our observation. The spectrum of the active pole was completely dominated by soft, blackbody-like emission with $kT_{\rm BB}\simeq 35$\,eV and showed no evidence for an optically thin hard X-ray emission characteristic of a cooling post-shock accretion column. Instead, we find constant, faint, optically thin plasma emission with a temperature of $\approx$ 1~keV at all orbital phases, which is uncorrelated with the soft X-ray emission, and we attribute this to shock-heated matter at the stagnation region between the two stars. Pronounced absorption dips punctuate the phase of active X-ray emission and they are associated with puffed-up material in the stagnation region. Using spectro-timing analysis of the absorption dips, the density profile and evolution of the clumpy flow were probed. We find evidence of a cyclotron harmonic from the southern pole (160 MG) near $\sim$2300\AA\ in the UV Grism spectrum. We also report minute-long quasi-periodic oscillations in the optical, UV, and X-ray light curves.}
{AR UMa represents an extreme case of soft, blobby accretion, with an effectively infinite soft excess, extending the magnetic field range over which magnetic accretion can be studied by a factor of two. It is one of the rare systems in which X-ray emission from a stream–magnetosphere interaction is detected.}
   \keywords{accretion, accretion disks – stars: novae, cataclysmic variables – X-rays: binaries – stars: magnetic fields –
stars: individual: AR UMa
               }
   \maketitle
     
\section{Introduction}
Magnetic cataclysmic variables (MCVs) are accreting compact binaries consisting of a white dwarf (WD) primary with a magnetic field $>$ 1~MG and a low-mass, main-sequence secondary. MCVs are divided into synchronized systems called polars and nonsynchronous systems called intermediate polars. The polars \citep{Cropper_Polars} harbor WDs with magnetic fields $>$10~MG, which force synchronization of the spin and binary orbit as well as prevent the formation of accretion reservoirs such as disks. The accretion flow in polars is instead understood to be in the form of streams and curtains. In stream accretion the matter that leaves the L1 follows a ballistic trajectory until the ram pressure of the stream ($\rho_{s}v_{\rm ff}(r)^{2}$) becomes comparable to the magnetic pressure ($\frac{B\small{(r)}^{2}}{8\pi}$), at which point the stream is threaded onto the field lines of the WD. This is demonstrated by Doppler tomographic studies of polars, which show components associated with both the ballistic and field-aligned part of the accretion flow \citep{Doppler_polars}. The field-aligned flow eventually plummets onto either one or both poles of the WD. Thus a substantial part of the accretion flow in polars is magnetically driven and so they form the ideal laboratories for understanding accretion physics at extreme magnetic fields. 

The canonical model of accretion in these systems envisages a supersonic accretion flow that undergoes a stand-off shock above the WD photosphere \citep{Lamb_Masters, King_Lasota_1979}. The shock-heated matter cools via cyclotron radiation and thermal bremsstrahlung ($\sim$20~keV). Roughly half of this shock energy is absorbed and reprocessed by the WD as soft X-ray blackbody radiation ($\sim$30~eV), so we expect $\frac{L_{\rm reprocessed}}{L_{\rm shock}} \approx 0.5$ (see \cite{2000Wu} for the hydrodynamics of the standing shock). However, several polars have shown deviations from this expectation, exhibiting soft X-ray luminosities far in excess of what is expected from reprocessing. This ``soft X-ray puzzle'' was first encountered in the observation of AM Her in a ``reversed'' accretion mode \citep{AMHer_rev} (in which the hard and soft emissions were found to be anticorrelated) and gained prominence during the \textit{ROSAT} era when several other polars were observed to show a soft excess (e.g., \cite{Beuermann_Softexcess_ROSATCVs, Ramsay_ROSATCVs}). Several theories were put forth to explain the soft excess and a consensus was developed around the idea of inhomogeneous or blobby accretion. First developed in the context of accreting neutron stars \citep{Blobs_NeutronStars}, the theory was adapted for magnetized WDs by \cite{Kupiers_Pringle} and further developed by \cite{1988FKL}. According to the blobby model, the accretion stream undergoes fragmentation at the stagnation region due to the development of exchange instabilities resulting in the formation of diamagnetic blobs. The denser blobs get tidally stretched, forming filaments of even higher densities. These filaments have a ram pressure much greater than the photospheric pressure at optical depths of $\tau \approx1$ on the undisturbed WD, such that they penetrate deep into the WD photosphere, burying the shock and fully thermalizing the shock energy, which eventually escapes as soft blackbody-like radiation. Depending on the degree of inhomogeneity in the accretion flow, all or a fraction of the shock energy can be released in the soft X-rays. According to this view, the soft X-ray excess is, in actuality, the absence of observable hard X-ray emission \citep[for a comprehensive review of the current understanding of buried shocks, please see][]{Wikram, 2003MHD_buriedshocks}.  For decades, since its conception, blobby accretion has been used to explain the apparent soft excess seen in several polars, including V1309 Ori \citep{V1309Ori}, QS Tel \citep{QSTel}, and the asynchronous system BY Cam \citep{BYCam}. Nevertheless, an understanding of the exchange instabilities leading to the formation of the blobs, the physics of their impact, its effect on the WD's atmosphere, among others, is still lacking. A possible correlation of the soft excess with magnetic field strength or, more specifically, the threading region was shown by \cite{Ramsay_ROSATCVs} using a sample of \textit{ROSAT} CVs. In the \textit{XMM-Newton} era, however, there is a paucity of research that directly addresses the accretion physics in polars, comprehensively. The only systematic study to do so to our knowledge is that of \cite{Ramsay_energybudget}, in which the authors reexamined the question of the energy budget of polars using an \textit{XMM-Newton} snapshot survey of 22 polars in their high states, along with the reanalysis of \textit{ROSAT} PSPC data using the then latest calibration. Contrary to the aforementioned \textit{ROSAT} studies, \cite{Ramsay_energybudget} argued against the existence of a soft excess in the majority of polars, instead they concluded that the soft excess is a rarity, only found in systems with high instantaneous accretion rates. Despite these rather equivocal findings and the ensuing ambiguity about the incidence and prevalence of a soft excess, the topic has received little attention in the last two decades. The aim of our paper is to investigate the nature of the accretion flow in the highest field polar, AR~UMa, and derive the energy budget of a polar at this magnetic extreme. For this, we utilized the first-ever high-state \textit{XMM-Newton} observation of the source.
 
AR~UMa was discovered as an extremely soft source in two \textit{Einstein} slew surveys and was optically identified with a known variable \citep{1994Remillard}. It is relatively close, at a distance of $98.62 \pm 0.52$~pc \citep{Bailer} and suffers very little Galactic extinction \citep{STIS_ARUMa}. Optical spectroscopy carried out during a 1991 low state revealed an M6-dwarf companion. A period of 1.932 hours, interpreted to be the binary's period, was estimated from photometric records. The two slew-IPC observations fortuitously taken during a high state revealed a very luminous soft blackbody of temperature $\approx$22 eV and led to the speculation of the system being a polar \citep{1994Remillard}. Historical and current photometric records show that AR~UMa is highly bimodal with historical lows of $V\sim16.5$ mag and rare highs reaching $V\sim13.5$ mag. This extremely low duty cycle is often attributed to its enormous magnetic field \citep{ARUMa_TESS}.  AR~UMa was the first high-field ($>$ 100MG) polar and remains the only one to exceed the 200~MG threshold. A magnetic field of 230 MG was estimated from the Zeeman splitting of Ly$\alpha$ in a low-state \textit{IUE} spectrum \citep{Schmidt_ARUMa_Zeeman_1996} and the low-state polarization properties. This enormous field was confirmed by \textit{HST} STIS spectroscopy carried out in a low state by \cite{STIS_ARUMa}, they also estimated a southern-pole strength of 160~MG consistent with the estimates from \cite{Field_weak}. Unlike prototypical polars, circular polarization of AR~UMa is undetectable in the high state and only a weak circular polarization of $V/I \approx +5\%$ is detected in its low state and is understood to result from photospheric dichroism of the magnetic WD \citep{1992Stockman}. The high soft X-ray luminosity ($3 \times 10^{33}$ erg s$^{-1}$ \citep{1994Remillard}) coupled with the lack of circular polarization indicated that the source may be an extreme case of soft X-ray excess \citep{Schmidt_ARUMa_Zeeman_1996}, quantitative evidence notwithstanding. In the X-rays, AR~UMa was observed by \textit{ASCA} in a low state and by \textit{RXTE} in a high state during which the 2-10 keV luminosity was estimated to be 1.8$\times$$10^{29}$ erg $s^{-1}$ and 9.4$\times$$10^{29}$ erg $s^{-1}$, respectively. This estimated high-state luminosity is far lower than typical high-state luminosities of most polars \citep{Szkody_EUVE_ASCA_RXTE_1999}. However, the \textit{EUVE} observation obtained simultaneously with \textit{RXTE} revealed the source to be EUV-bright with the light curve indicating the primary accreting pole to be southern, that is, below the plane from the observer’s perspective, consistent with the geometry derived by \cite{Schmidt_ARUMa_Geometry_1999}. Though the source has been subject to the aforementioned extensive spectroscopic, photometric, and spectro-polarimetric campaigns, it has evaded a high-state observation in the soft X-rays since the time of its serendipitous discovery. AR~UMa entered a protracted low state in 2008 lasting more than a decade. With the intention of capturing the high state of the source in the soft X-rays, AR~UMa was made part of an optical monitoring campaign with the help of \textit{AAVSO} \citep{AAVSO} and a Target of Opportunity (ToO) program with \textit{XMM-Newton}. In 2023, when the source reached an optical magnitude <15.8 mag the \textit{XMM-Newton} observation was triggered. Concurrently with the X-ray observation we obtained optical photometric data using the 70-cm reflector at the AIP. The paper is organized as follows. Section~\ref{section:obs_red} details the observations and data reduction. Section~\ref{section:timing} and Sect.~\ref{section:spectra} describe the temporal and spectral analyses, respectively. Section~\ref{results_discussion} presents our results and discussion, and Sect.~\ref{conclusion} summarizes our main conclusions.

\begin{table}
\centering
\footnotesize
\setlength{\tabcolsep}{3pt}
\caption{Observation log for XMM-Newton (Obs ID: 0884870101; PI: Axel Schwope).}
\begin{tabular}{cccc}
\noalign{\smallskip}
\hline\hline
Instrument & Exposure ID & Exposure length & Mean count rate \\
 & & (s) & (cts s$^{-1}$) \\
\noalign{\smallskip}
\hline
\noalign{\smallskip}
EPIC MOS-1  & S001 + U003 & $26647 + 7942$ & $1.1 \pm 0.1$ \\
EPIC MOS-2  & S002        & $43005$        & $1.2 \pm 0.1$ \\
EPIC pn     & S003        & $42539$        & $14.3 \pm 0.7$ \\
RGS-1       & S004        & $43313$        & $0.10 \pm 0.002$ \\
RGS-2       & S005        & $43308$        & $0.09 \pm 0.002$ \\
OM-UVW1     & S006-12     & $\sim 4400 \times 7$ & $66.8 \pm 0.4$ \\
OM-UV Grism & S013-16     & $\sim 2039 \times 4$ & $2.31 \pm 0.1$ \\
\noalign{\smallskip}
\hline
\end{tabular}
\label{tab:log}
\end{table}

\section{Observations and data reduction} \label{section:obs_red}
\subsection{XMM-Newton observations} 
AR~UMa was observed by \textit{XMM-Newton} as a ToO on $9^{th}$ May 2023 during a rare high state with V-mag reaching $\sim 13.8$. All EPIC cameras \citep{PN, MOS} were operated in the small window mode using the medium filter. The EPIC pn and EPIC MOS-2 observations covered $\sim$ 6.2 orbital cycles amassing a total exposure time of $\sim 43$ ks whereas, EPIC MOS-1 missed most of the fifth cycle and has a shorter exposure time of $\sim 36$ks.  The reflection grating spectrometer (RGS) \citep[][]{RGS} was used simultaneously with the EPIC instruments to obtain high-resolution soft X-ray spectroscopic measurements. The optical monitor (OM) \citep{OM} was utilized two-fold, namely, in fast mode using the UVW1 filter ($\lambda_{\rm eff} = 2910 \AA$) for seven sub-exposures with a combined exposure time of $\sim$ 30 ks and using the UV-Grism ($1800\ \AA$ -- $3600 \AA$) for another set of sub-exposures totaling $\sim$ 7.8 ks, obtaining low-resolution UV spectra. The log of the observation is shown in Tab.~\ref{tab:log}. The data were reduced using the science analysis system, \textsc{sas v.19.1}. Parts of the EPIC observation were affected by background flaring. To eliminate these instances from analysis, Good Time Intervals (GTIs) were constructed from which science products were extracted. The light curves were extracted with the maximum possible resolution allowed by the small window mode, that is, 5.7~ms for EPIC pn and 0.3~s for EPIC MOS-1\&2. The source was unprecedentedly bright during the time of the observation leading to considerable pile-up in EPIC cameras. The pile-up was assessed using \textsc{``epatplot''}, which compares the observed single and double pattern fractions against model predictions. To mitigate pile-up, the core of the point spread function (PSF) was excised with an inner radius of 12 arcsec for EPIC pn and 18.6 arcsec for EPIC MOS-1\&2, and spectra were extracted from an annular region with an outer radius of 35 and 40 arcsec respectively. After excision, the observed pattern fractions were consistent with model predictions. The RGS-1 \& 2 data were extracted using the standard tasks and their respective orders were combined using \textsc{``rgscombine''}. The OM fast-mode data were reduced using \textsc{``omfchain''} with a time resolution of 1 second and the count rates were converted to flux units using the conversion factor  $4.76 \times 10^{-16}$ erg cm$^{-2}$ s$^{-1}$ $\AA$$^{-1}$ cts$^{-1}$ given in the \textit{XMM-Newton} documentation \footnote{\href{https://www.cosmos.esa.int/web/xmm-newton/sas-watchout-uvflux}{OM count rate to flux conversion.}}. The UV-Grism data were reduced using \textsc{``omgchain''} obtaining four different exposures of UV-Grism spectra which were also coadded to create an average spectrum.
\subsection{Optical photometry}
\begin{figure*}
\centering
        \includegraphics[width=0.72\textwidth]{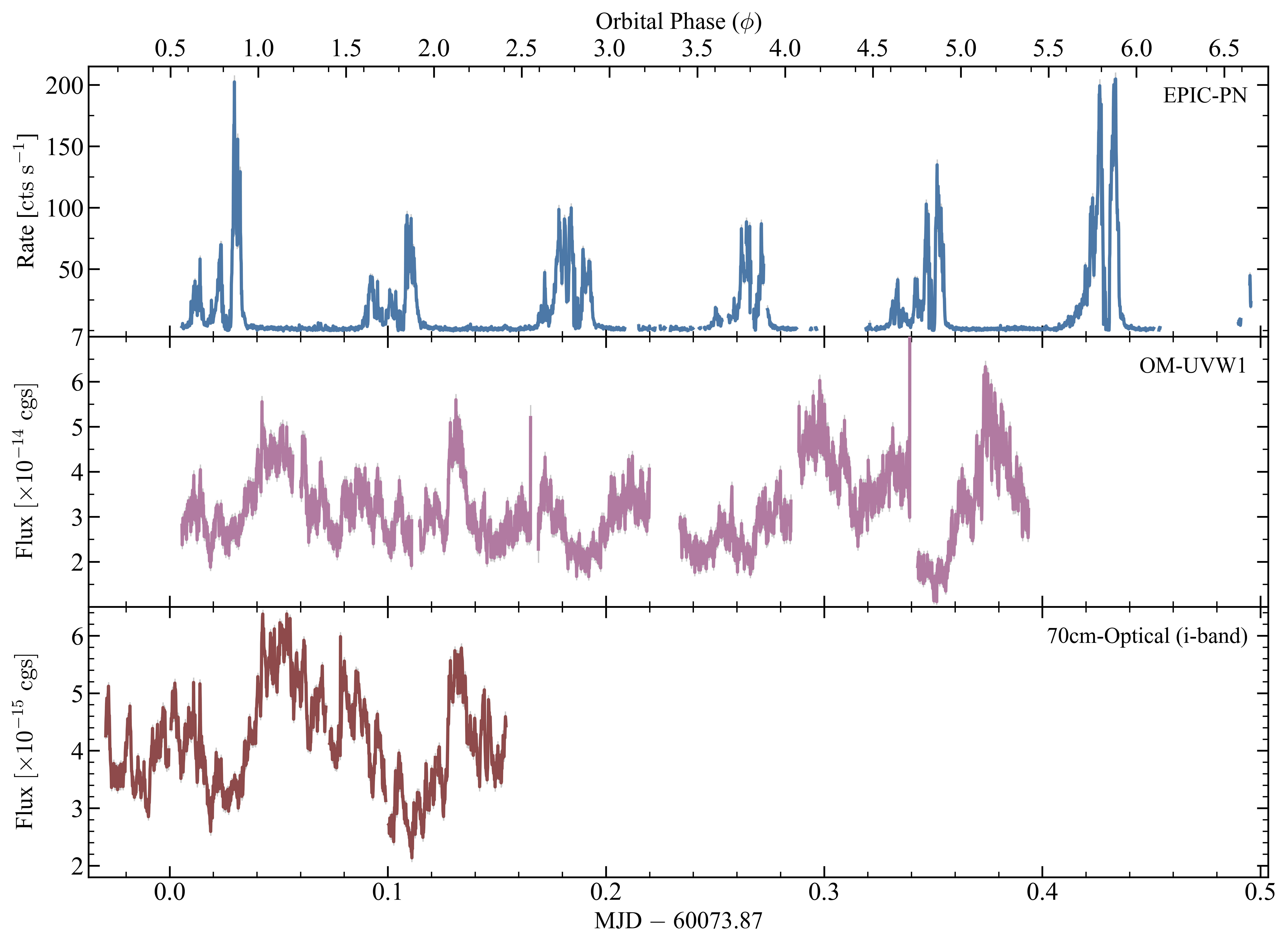}
    \caption{Top down: \textit{XMM-Newton} EPIC pn light curve of AR~UMa in the energy range 0.15-15 keV, OM-UVW1 fast light curve, and optical i-band light curve obtained from the 70-cm reflector at AIP. All light curves are binned to 10s. The gaps in the EPIC pn light curve result from background flare filtering, whereas those in the OM-UVW1 light curve correspond to the intervals between the seven separate OM-fast exposures.
}    
    \label{fig:xmm_lcs}
\end{figure*}

AR UMa was observed with the 70 cm reflector located at the Babelsberg site of the Leibniz-Institut für Astrophysik Potsdam (AIP). The telescope is equipped with a back-illuminated sCMOS camera that allows fast readout at rather low noise (Kepler 400). Observations were performed through an SDSS i-filter on the night 2023 May 9 for a total of 4.5 hours under good weather conditions with some thin cirrus only at the beginning of the observations. Differential photometry on dark-subtracted and flatfield-corrected images using ESO/MIDAS was performed to determine the magnitude differences of AR UMa with respect to a comparison star, the brightness of which was recorded on the same frame(s). The star used is Gaia DR3 783918259794939776 at RA: 11h15m50.472s, and Dec: +42\degr\,55'11.20". The SDSS lists mean $ugriz$ magnitudes for this star of 14.95, 13.07, 12.31, 12.16, and 13.14 mag, respectively, and we used the given $i$-band value to calibrate the magnitude scale. A total of 1423 valid frames were obtained covering 2.28 binary cycles starting at cycle 119302.12 according to \cite{Schmidt_ARUMa_Geometry_1999}. The resulting light curve is shown in Fig.~\ref{fig:xmm_lcs}.
\section{Timing analyses} \label{section:timing}

\subsection{X-ray light curves}
\begin{figure}
    \centering
    \includegraphics[width=0.7\columnwidth]{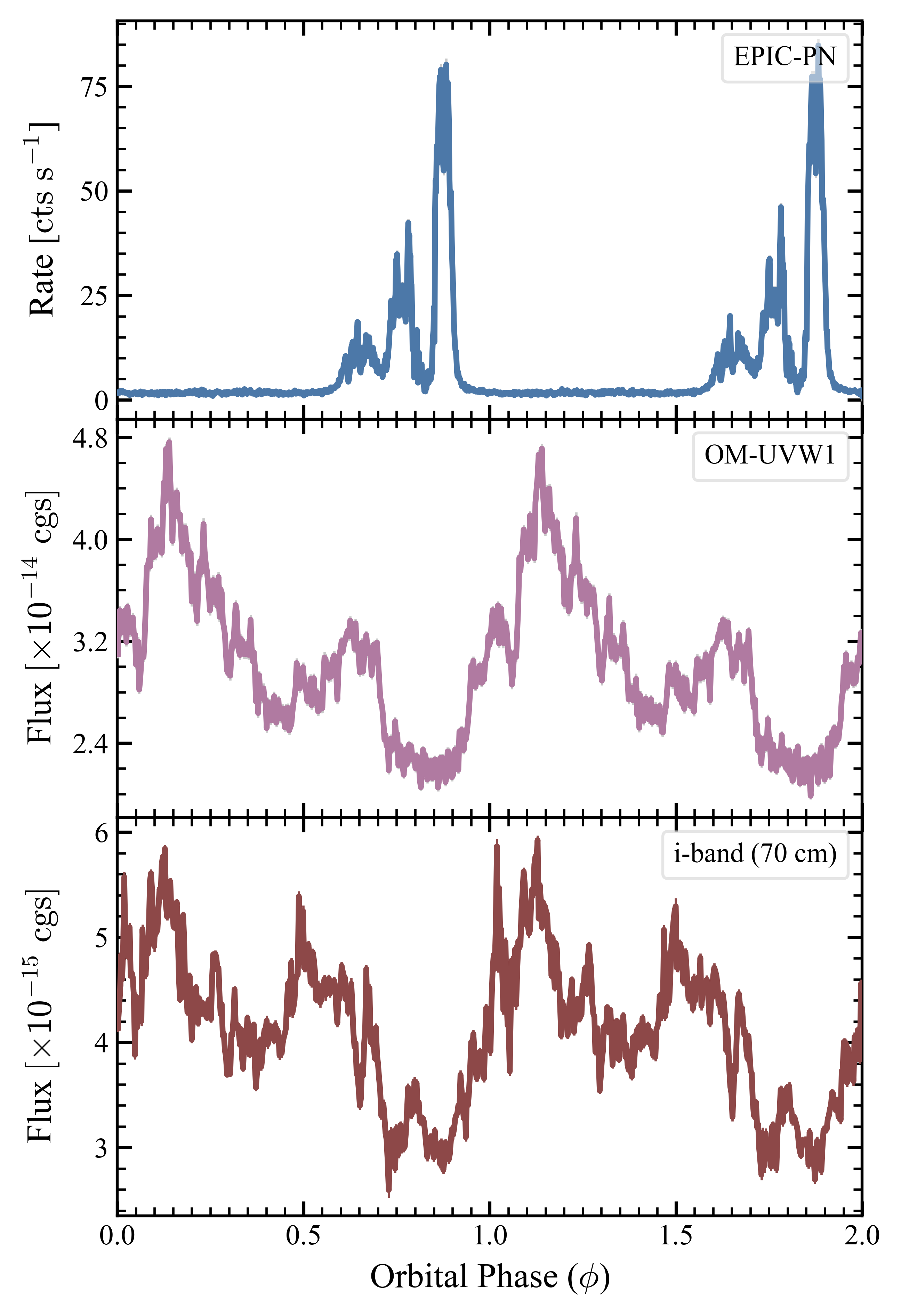}
    \caption{Top down: Phase-folded \textit{XMM-Newton} EPIC pn (0.15-15 keV), OM-UVW1, and i-band (70-cm) light curves of AR~UMa, with a binning of 10s. Phase 0 corresponds to the inferior conjunction of the secondary.}
    \label{fig:phase_fold}
\end{figure}

The \textit{XMM-Newton} X-ray light curve obtained with EPIC pn in the energy range 0.15-15 keV spanning the entire observation ($\approx$ 6.5 orbital cycles) is shown in Fig.~\ref{fig:xmm_lcs}. 
The light curve shows dramatic variations in its morphology from cycle to cycle. To characterize the prominent features of the light curve we phase-folded it to the orbital ephemeris from \cite{Schmidt_ARUMa_Geometry_1999}, where phase-zero corresponds to the inferior conjunction of the secondary star. We note that extrapolating this ephemeris to our observing epoch introduces an accumulated phase uncertainty of $\Delta\phi=\pm0.18$. The EPIC pn (0.15-15 keV) light curve phase-folded and binned to 10~s is shown in the top panel of Fig.~\ref{fig:phase_fold}. The morphology of the light curve is similar to systems that undergo a self-eclipse of the X-ray emitting pole, for example, QS Tel \citep{QSTel}, EF Eri \citep{StreamDip_EFEri}, VV Pup \citep{VVpup_Soft}, and V1309 Ori \citep{V1309Ori}. The striking feature in the light curve is a bright interval with peak count rate reaching a maximum of $84.7 \pm1.5\ $ cts s$^{-1}$ at 10~s resolution ($96.8 \pm 5.2$ cts s$^{-1}$ at 1~s resolution). We define the bright phase, that is, phase $\phi \simeq 0.58-0.93$, as the first and last instance of the light curve at which the mean count rate exceeds three times the count rate of faint interval spanning the phase range $\phi \simeq 0.2-0.5$. Using the duration and midpoint of the bright phase, $\Delta\phi = 0.35$ and  $\phi_{\rm mid} = 0.755$ respectively, and the system inclination $i \approx 70^\circ$ \citep{ARUMa_TESS}, we can estimate the geometry of the primary accreting pole. The azimuth follows directly from the bright-phase midpoint as $\psi = 360^\circ \times (1 - \phi_{\rm mid}) = 88.2^\circ$. Since the duration of the bright phase is $<$ 0.5 we locate the accreting pole in the southern hemisphere, the colatitude ($\beta$) of which is given by $\cos(\pi\Delta\phi) = -\cot i \cot\beta$ \citep{WorpelSH}; substituting the respective values of AR UMa yields $\beta$ $\simeq$ 142$^\circ$.

High-frequency variations that cannot be attributed to mere flickering are evident in the bright phase and are investigated in detail in Sect.~\ref{subsection:QPO}. This interval also shows prominent flares characteristic of blobby accretion \citep[e.g., V1309 Ori or AM Her, ][]{V1309Ori, AMHer_Axel}. In an inhomogeneous flow flares occur as a result of blobs impinging on the WD. However, unlike in V1309 Ori \citep{V1309Ori} in which the individual accretion events are uniquely identifiable, in AR~UMa the light curve shows soft flares superposed over an undulating bright emission and the light curve cannot entirely be resolved into individual flaring events. This behavior is reminiscent of the light curve of the soft polar AI Tri in which the emission is understood to result from a superposition of individual accretion events \citep{AI_Tri}. However, there are a few flaring events or clusters thereof, that are resolvable in the light curve and are characterized in detail in Appendix \ref{blob_appendix}.

The bright-phase emission is interrupted by two prominent absorption features. The primary event, occurring at orbital phase $\phi \sim 0.80-0.845$, is sharp and broad ($\Delta\phi=0.045$). Although it is present in all the observed cycles, it is not phase-locked. Additionally, a secondary, narrow and shallow absorption dip is seen near $\phi\simeq0.7$ which is sporadic and weak, and completely absent in some cycles (Fig.~\ref{fig:xmm_lcs}). These twin dips seen in polars are generally understood to be resulting from the line-of-sight obscuration of the emission region caused by the accretion flow, with the primary and secondary dip caused by the leading (upstream) and trailing (falling) edge of the stream, respectively \citep{StreamDip_EFEri}. However, this explanation cannot be applied to AR UMa during our observations, because AR UMa seemed to have just one active pole, which is southern (below the plane from the observer's perspective) and, hence, we do not expect to see a stream dip at all, unless the obscuring structure is further out in the stagnation region and has a large geometrical cross-section (refer Sect.~\ref{stream_geometry}). There is significant activity and substructure within the stream dip indicating density fluctuations in obscuring structure, suggestive of a clumpy flow. The long observation covering six orbital cycles provides the opportunity to directly probe the cycle-to-cycle variation of the inhomogeneous accretion flow and is presented in Sect.~\ref{subsubsection:stream-dip}.

\begin{figure}
    \centering
    \includegraphics[width=0.72\columnwidth]{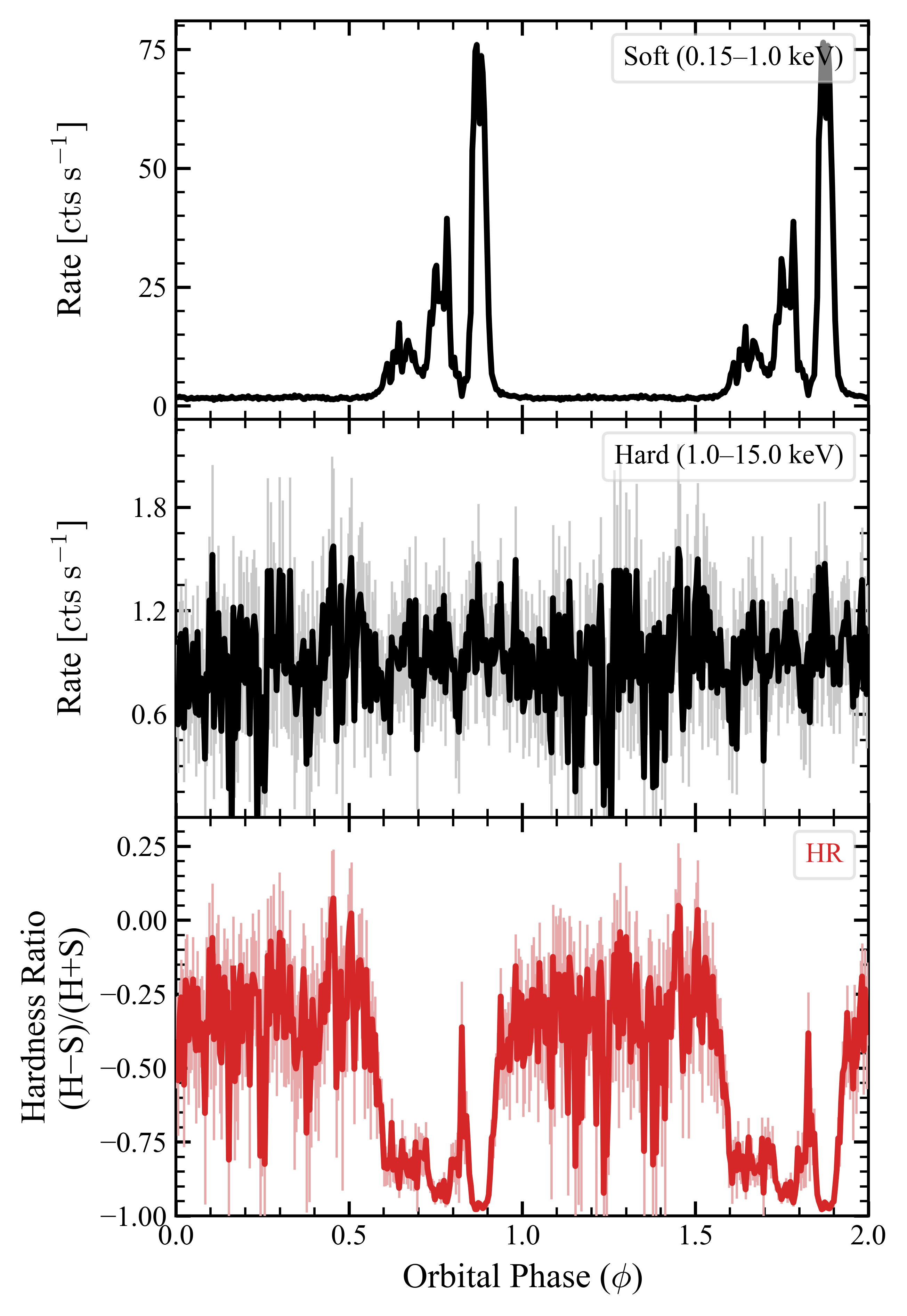}
    \caption{Energy-resolved \textit{XMM-Newton} EPIC pn light curve of AR~UMa binned at 30s. The soft and hard bands correspond to 0.15-1 keV and 1-15 keV, respectively. The hardness ratio is calculated as $\frac{H-S}{H+S}$ where H is the count rate in the hard band and S is the count-rate in the soft band.}
    \label{fig:pn_energyresolved}
\end{figure}

The unambiguous soft excess in AR~UMa is perhaps best captured by the energy-resolved light curve shown in Fig.~\ref{fig:pn_energyresolved} along with the hardness-ratio curve ($HR = \frac{H-S}{H+S}$) calculated using the soft band, 0.15-1~keV, and hard band, 1-15~keV. \text{The} bulk of the emission emanating from the source is clearly below 1~keV. During the bright phase ($\phi$ = 0.58-0.93) the soft-band mean count rate is $20.7\pm0.05$ cts s$^{-1}$ and the hard-band mean is $0.94\pm0.04$ cts s$^{-1}$ whereas in the faint phase ($\phi$ = 0.2-0.5) the mean count rate is $1.72\pm0.02$ cts s$^{-1}$ and $0.91\pm0.05$ cts s$^{-1}$, in the soft and hard bands, respectively. Interestingly, there are no photons above 1~keV that can be exclusively attributed to the bright-phase emission. The hard X-ray light curve is very faint and does not show orbital phase-dependent modulation, resembling the high-state \textit{RXTE} light curve from \cite{Szkody_EUVE_ASCA_RXTE_1999}. Whereas, the soft X-ray (0.15-1.0 keV) is bright and heavily modulated on the orbital period and is identical to the high-state \textit{EUVE} light curve reported by \cite{Szkody_EUVE_ASCA_RXTE_1999}, indicating the EUV and the soft X-ray emission arise from the same location and is consistent with the picture of buried shocks. As in the EUV regime, the southern pole dominates the soft emission with the northern pole contributing little light, if at all. The absorption dips which are completely absent in the hard X-ray light curve are pronounced in the softer energies, during which the HR climbs to $-0.71\pm0.03$ from an average of $\sim -0.91$ during the periods of heightened emission, this energy-dependency of the absorption attests to its photoelectric nature.


\subsection{Optical and UV light curves}
The OM-UVW1 UV light curve and the \textit{70-}cm $i$-band light curve binned to 10~s are shown in Fig.~\ref{fig:phase_fold}. Both light curves show a similar asymmetric double-humped profile, unlike the low-state optical light curve which shows a symmetric profile \citep{Szkody_EUVE_ASCA_RXTE_1999, ARUMa_TESS}. The optical light curve shows an additional period of excess emission between phases $\phi\sim 0.78-0.85$. Both light curves show strong stochastic flaring and signatures of quasi-periodic variations (see Sect.~\ref{subsection:QPO}.). The general profile of the light curves, characterized by unequal double-peaks, is interpreted as resulting from the varying projection of the accretion stream. During the high state the irradiated surface of the secondary may also contribute to the optical and UV light. The optical and UV emission is antiphased to the X-ray emission and implies negligible contribution from the accretion spot itself. The additional emission seen in optical between $\phi\sim 0.78-0.85$ may emanate from the spot but is hard to substantiate due to the highly instationary nature of the accretion process.

\subsection{High-frequency variations}\label{subsection:QPO}
\begin{figure}[h] 
    \centering
        \centering
        \includegraphics[width=0.43\textwidth]{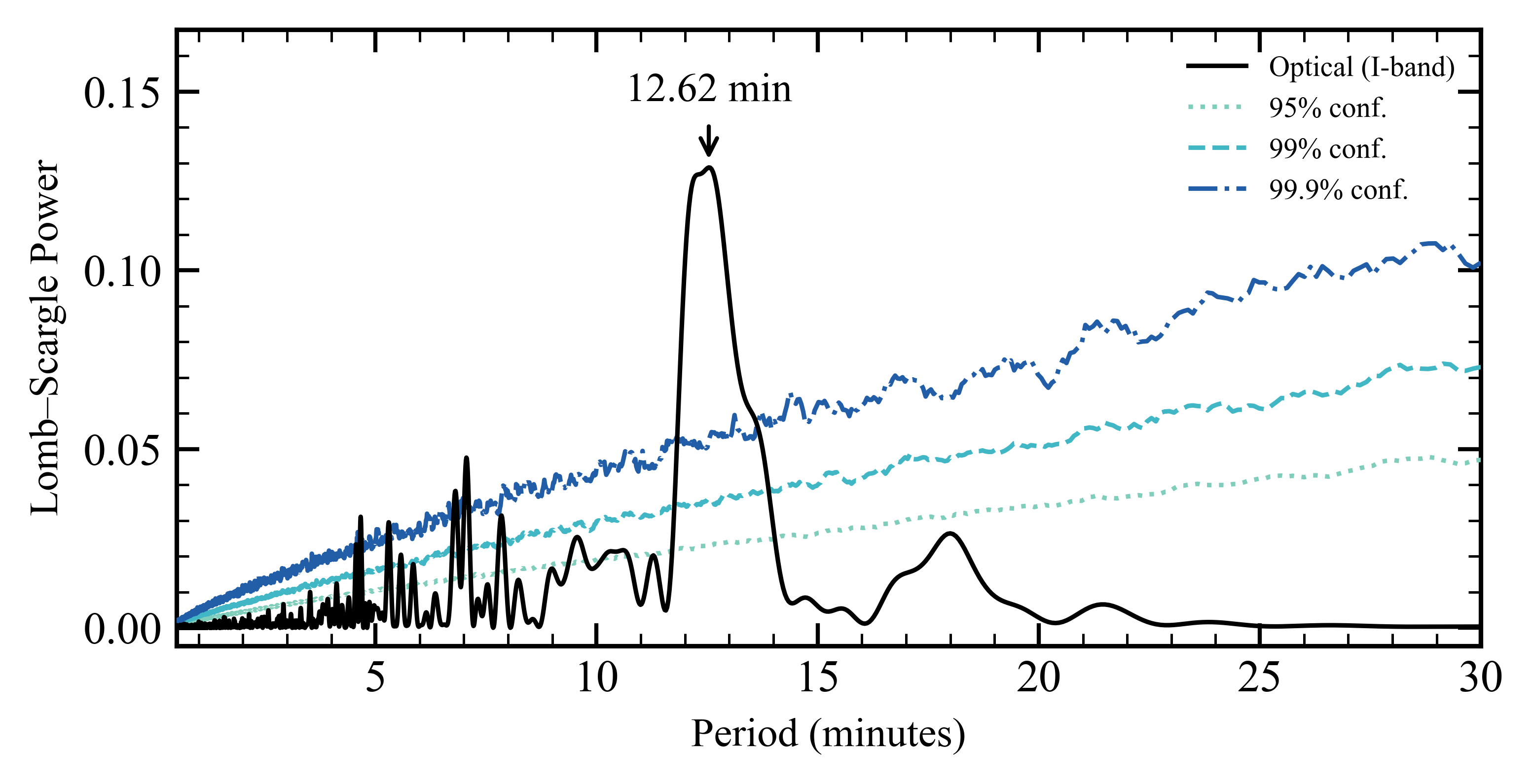}
          \vspace{-0.15cm} 
        \includegraphics[width=0.43\textwidth]{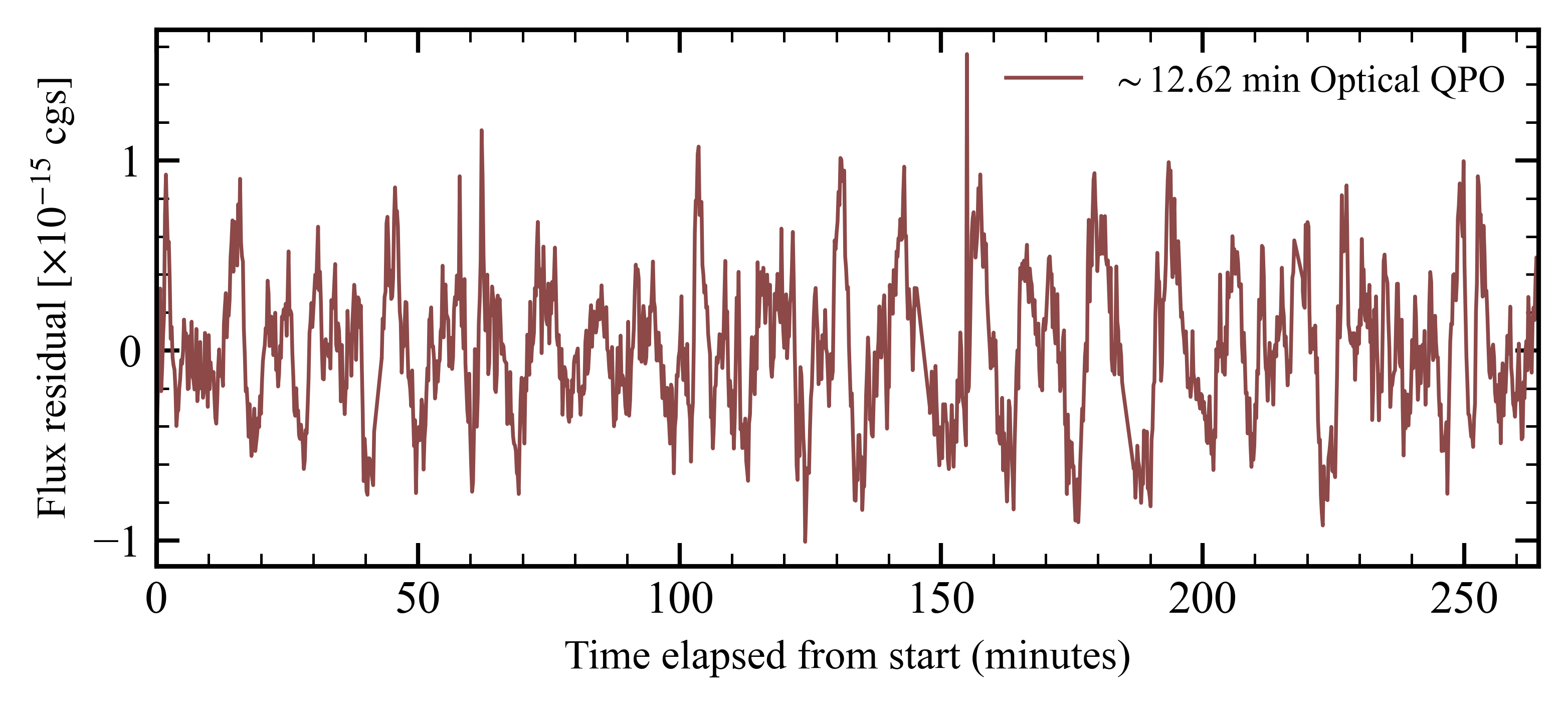}
    \vspace{-0.15cm} 
        \includegraphics[width=0.43\textwidth]{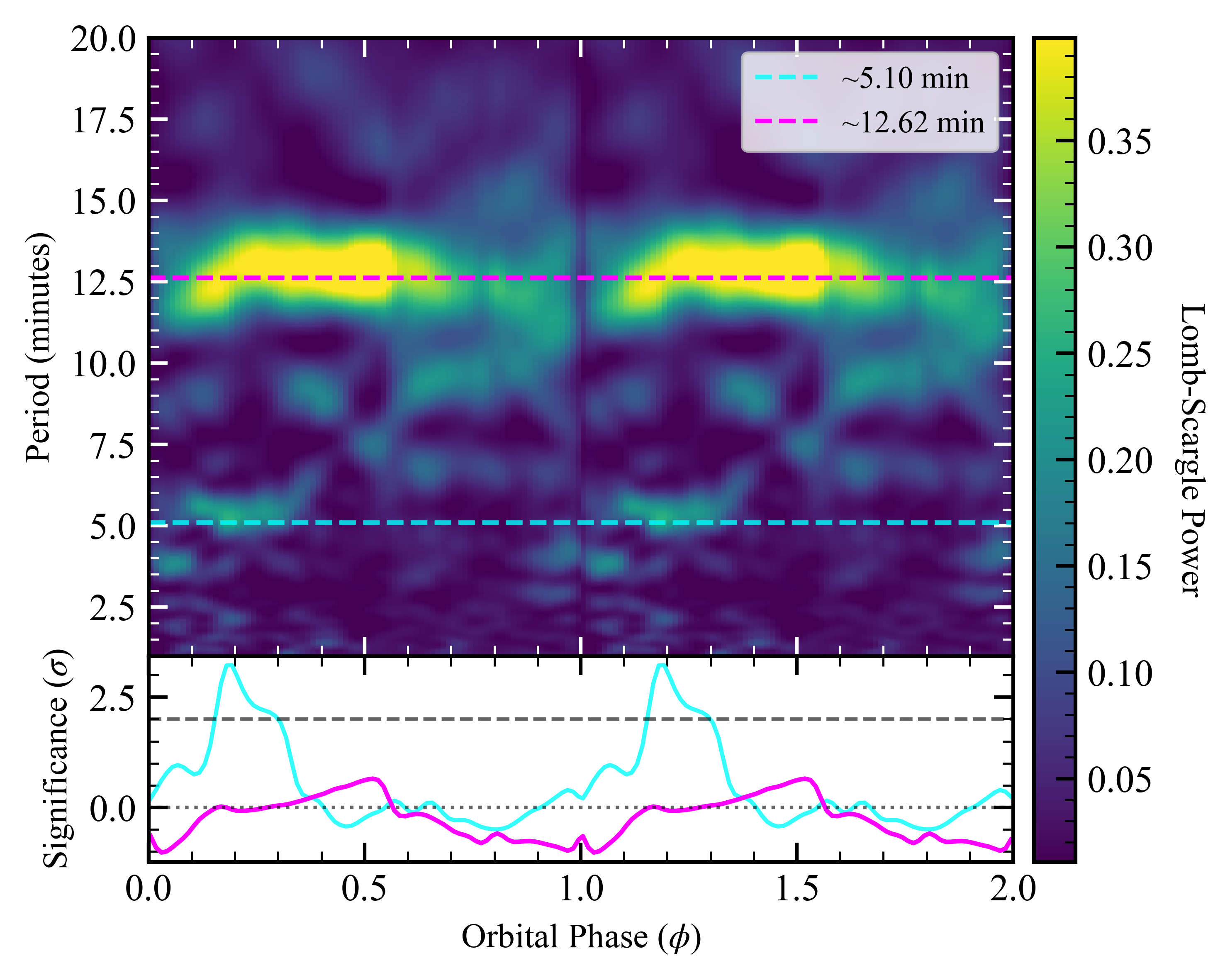}
    \caption{(Top) Lomb–Scargle periodogram of the detrended optical ($I$-band) light curve. The solid black line shows the power spectrum after removing the long-term trend. (Middle) Optical light curve residual showing QPO. (Bottom) Dynamical power-spectra of the optical light curve showing possible phase-dependency of the $\approx$ 12.6 min QPO and an intermittent QPO with $\approx$ 5 min period.}
    \label{fig:opt_qpo}
\end{figure}

\begin{figure}[h] 
    \centering
        \includegraphics[width=0.43\textwidth]{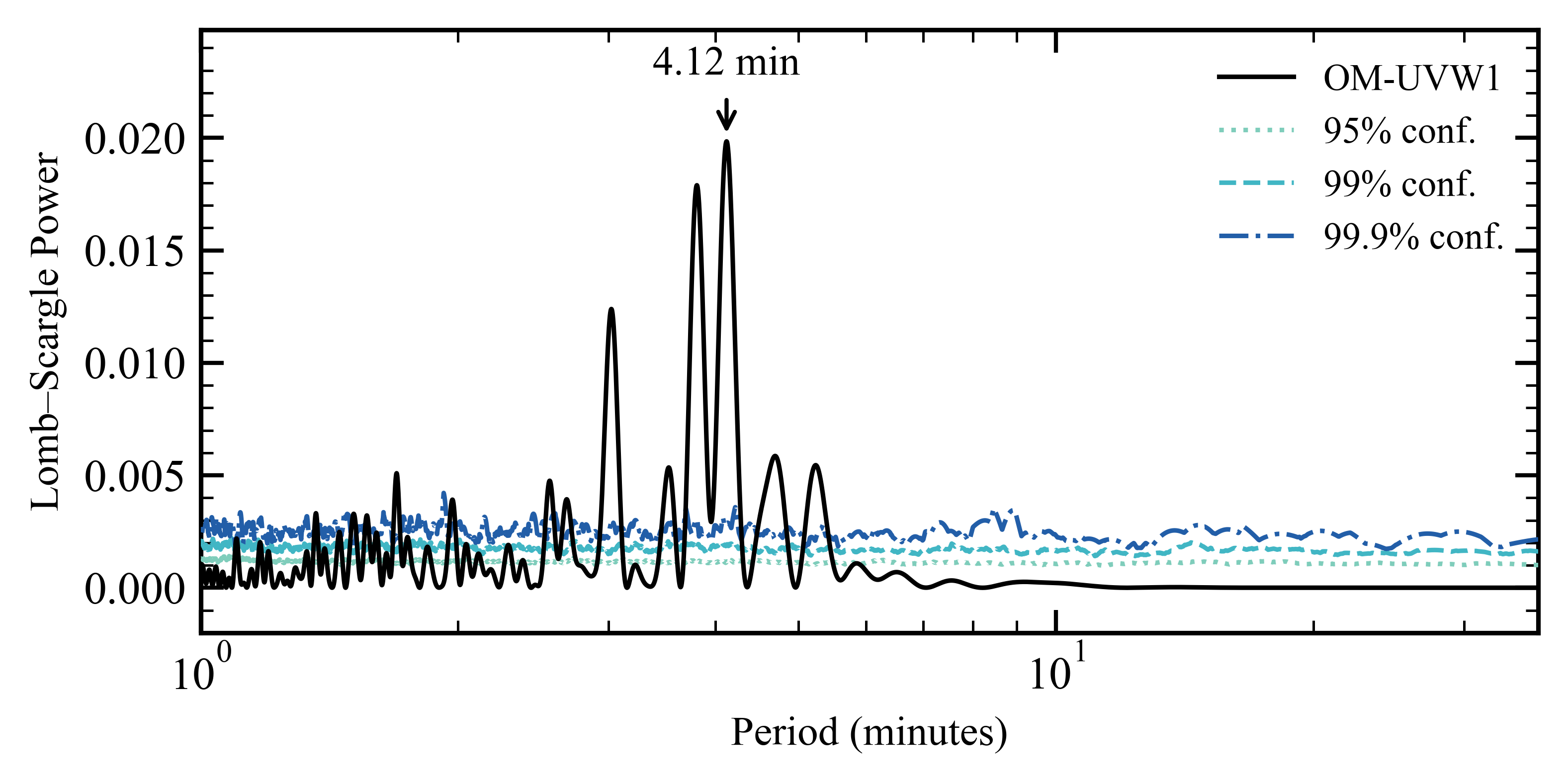}
          \vspace{-0.15cm} 
        \includegraphics[width=0.43\textwidth]{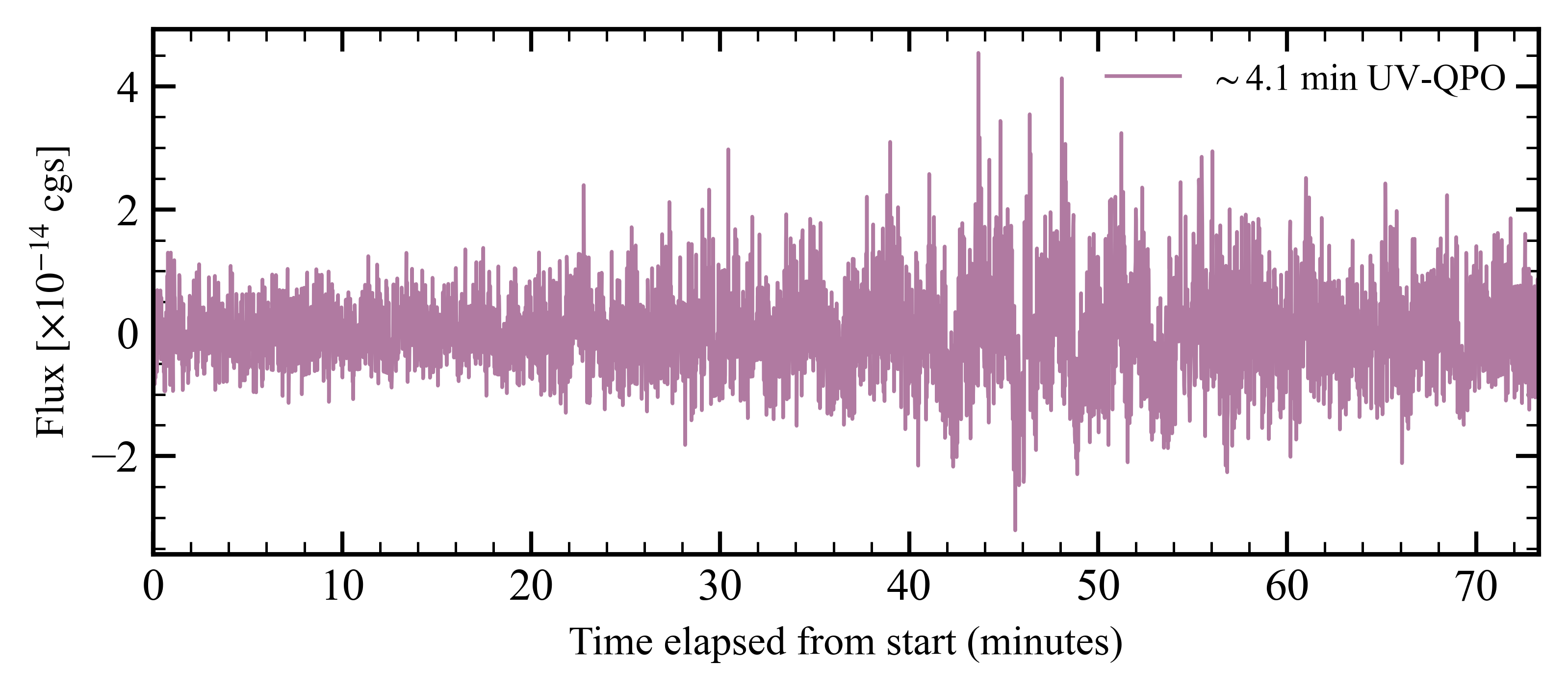}
    \vspace{-0.15cm} 
        \includegraphics[width=0.43\textwidth]{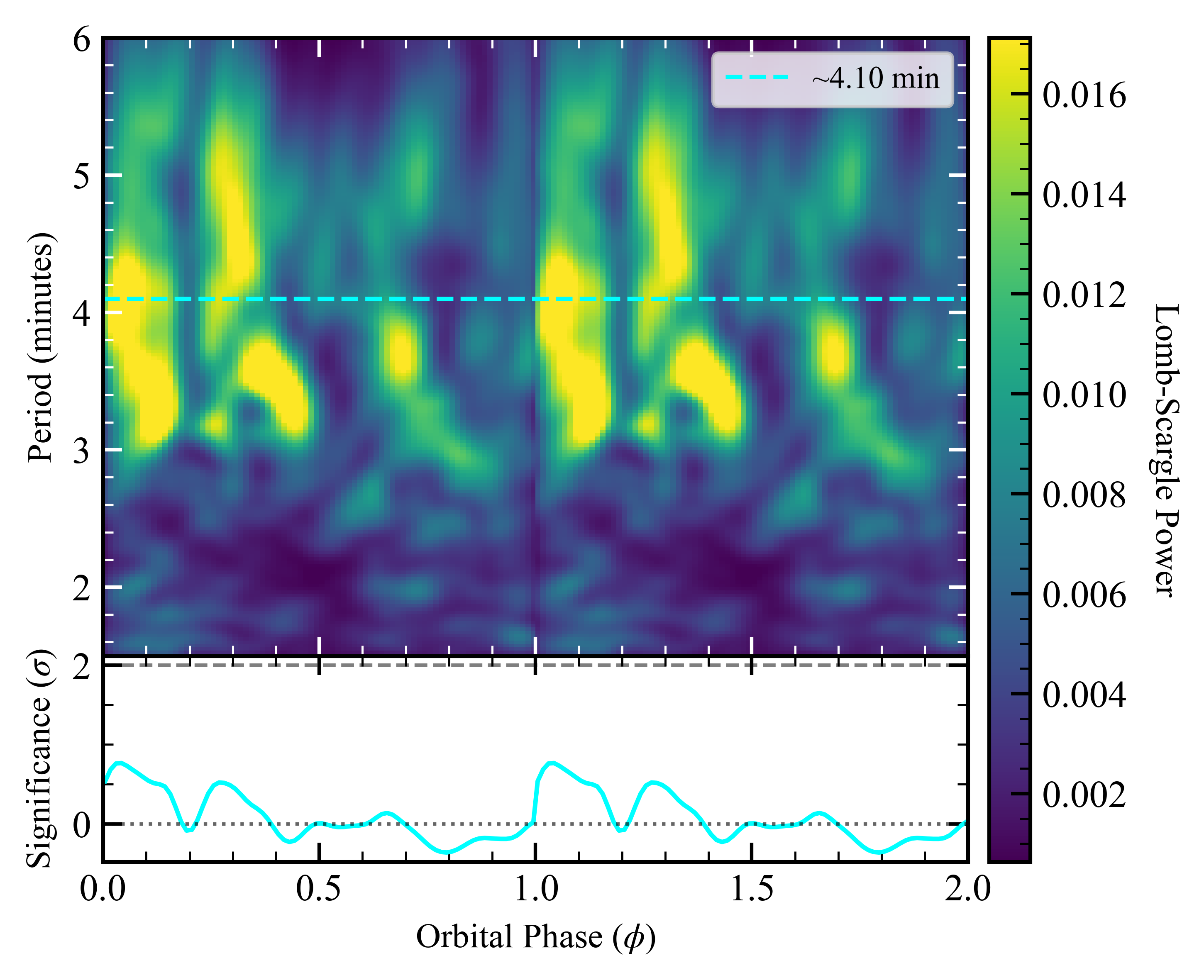}
    \caption{(Top)  Lomb–Scargle periodogram of the detrended OM UVW1 light curve. The solid black line shows the power spectrum after removing the long-term trend. (Middle) OM-UVW1 (S006) light curve residual showing QPO. (Bottom) Dynamical power-spectra of the merged UV light curve showing possible phase-dependency of the QPO.}
    \label{fig:uv_qpo}
\end{figure}

Quasi-periodic oscillations (QPOs) akin to second and/or minute-long QPOs which have been reported in several soft polars; AI Tri \citep{AI_Tri}, EF Eri \citep{EF_Eri_minQPO, EF_Eri_QPO}, AN UMa and V834 Cen \citep{AN_UMa_V834Cen}, among others, are clearly discernible in the multiwavelength light curves of AR~UMa, as demonstrated for instance by high-resolution light curves in Fig.~\ref{fig:opt_qpo}, and Fig.~\ref{fig:uv_qpo}. In order to determine the dominant frequencies underlying such short-period oscillations, we computed Lomb-Scargle periodogram \citep{Lombscargle} on the detrended residuals of highest available cadence optical (10~s), UV (1~s), and X-ray (0.15-1 keV ; 0.5~s) light curves. The periodograms show broad maxima or forest of signals typically associated with QPOs but could also be associated with flickering. To robustly assess the significance of these short-period variations against flickering driven by instationary mass-transfer, we followed the Monte Carlo method of \cite{2001Benloch}. For each wavelength, a power-law noise spectrum was fit to the respective observed periodogram. Using \cite{Timmer_Konnig} algorithm 10000 synthetic light curves with the same mean, variance, and power-law index as the residuals were simulated on the same time grid as the observations and their periodograms were computed. The 95\%, 99\%, and 99.9\% percentiles of the resulting power distribution at each frequency define the confidence contours plotted against the periodogram of the observed data, Fig.~\ref{fig:opt_qpo}, \ref{fig:uv_qpo}, and \ref{fig:xray_qpo}. In the power spectra of the optical (Fig.~\ref{fig:opt_qpo}), UV (Fig.~\ref{fig:uv_qpo}), and X-ray (Fig.~\ref{fig:xray_qpo}) we find dominant signals at 12.6~min, 4.12~min, and 14.5~min respectively, that exceed the 99.9\% significance threshold. To evaluate the orbital-phase dependency of the QPOs we constructed sliding-window Lomb-Scargle spectrograms with band-specific window width such that each window encompasses at least five cycles of the expected QPO. The windows were stepped at a cadence of 2~min for the optical data and 20~s for both UV and X-ray, and Lomb-Scargle power was estimated for the respective QPO frequency. The significance of the QPO frequency in each window was estimated using the median and standard deviation of the Lomb-Scargle power of the frequency computed across all sliding-windows. Each window center was then mapped to an orbital phase using the ephemeris and both the raw power and the significance were phase-folded into 80 phase bins and smoothed. The resulting dynamic power spectra of the optical, UV, and X-ray data are shown in Fig. ~\ref{fig:opt_qpo}, \ref{fig:uv_qpo}, and \ref{fig:xray_qpo}, respectively. 
The optical dynamic power spectrum reveals the $\sim$ 12.6 min QPO to dominate in the phase-range $\phi\sim 0.0-0.6$, additionally another transient signal at $\approx$ 5.1 min between $\phi\sim 0.0-0.3$ is also detected. We note that this $\sim$ 5.1 min period is equivalent (within errors) to the break-frequency detected by \cite{ARUMa_TESS} in the PSD of TESS data of AR UMa. The 5.1 min QPO is found to show a much higher significance than the 12.62 min QPO. The dynamic power spectra of the UV data reveal the UV QPO at 4.1 min to appear in the same phase range as the optical 12.6 min QPO albeit with very low significance. The 14.5 min X-ray QPO emanates in the X-ray bright phase and is likely to result from the absorption events seen in this interval.

\begin{figure}[h] 
    \centering
        \includegraphics[width=0.43\textwidth]{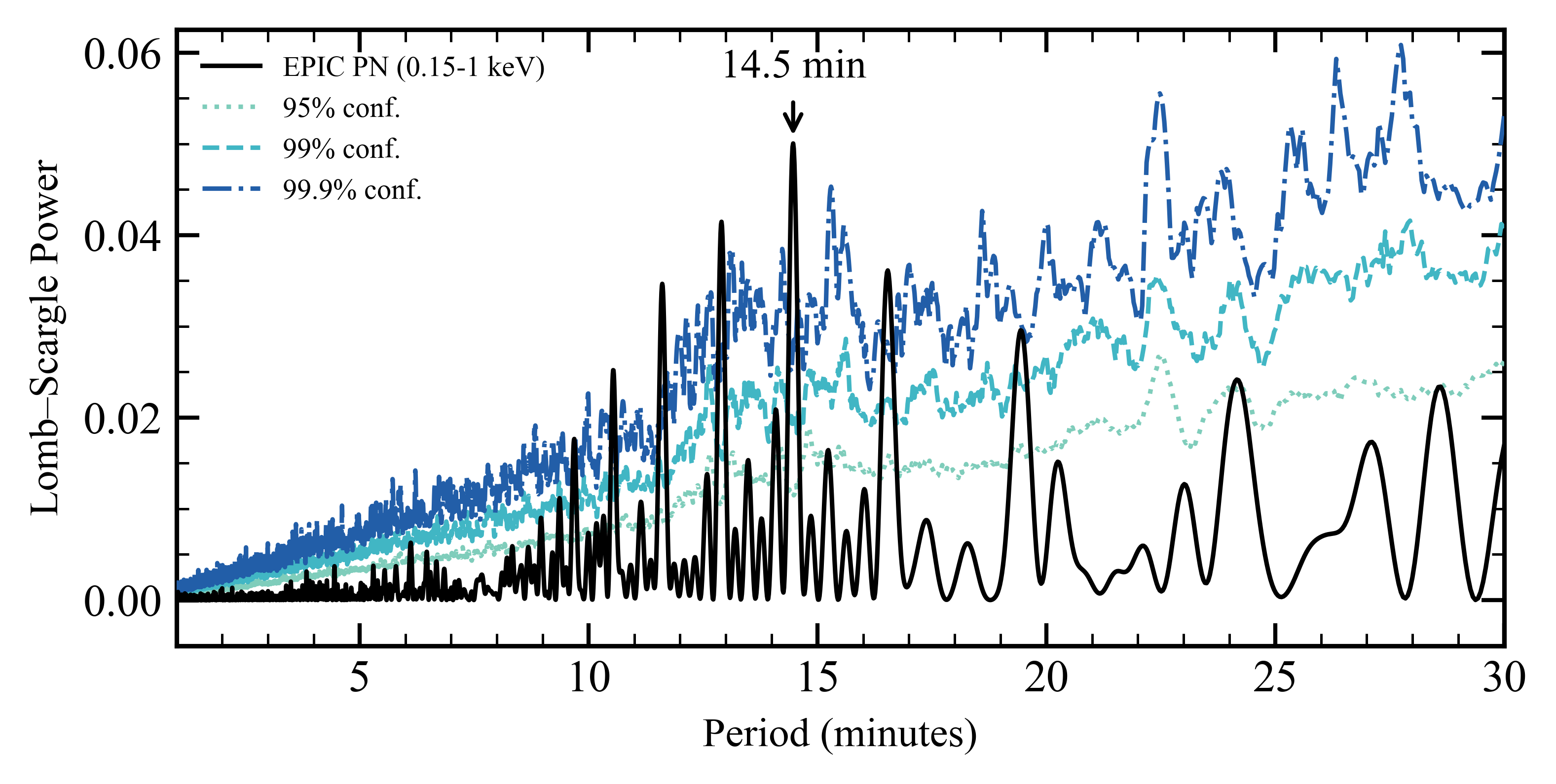}
          \vspace{-0.15cm} 
        \includegraphics[width=0.45\textwidth]{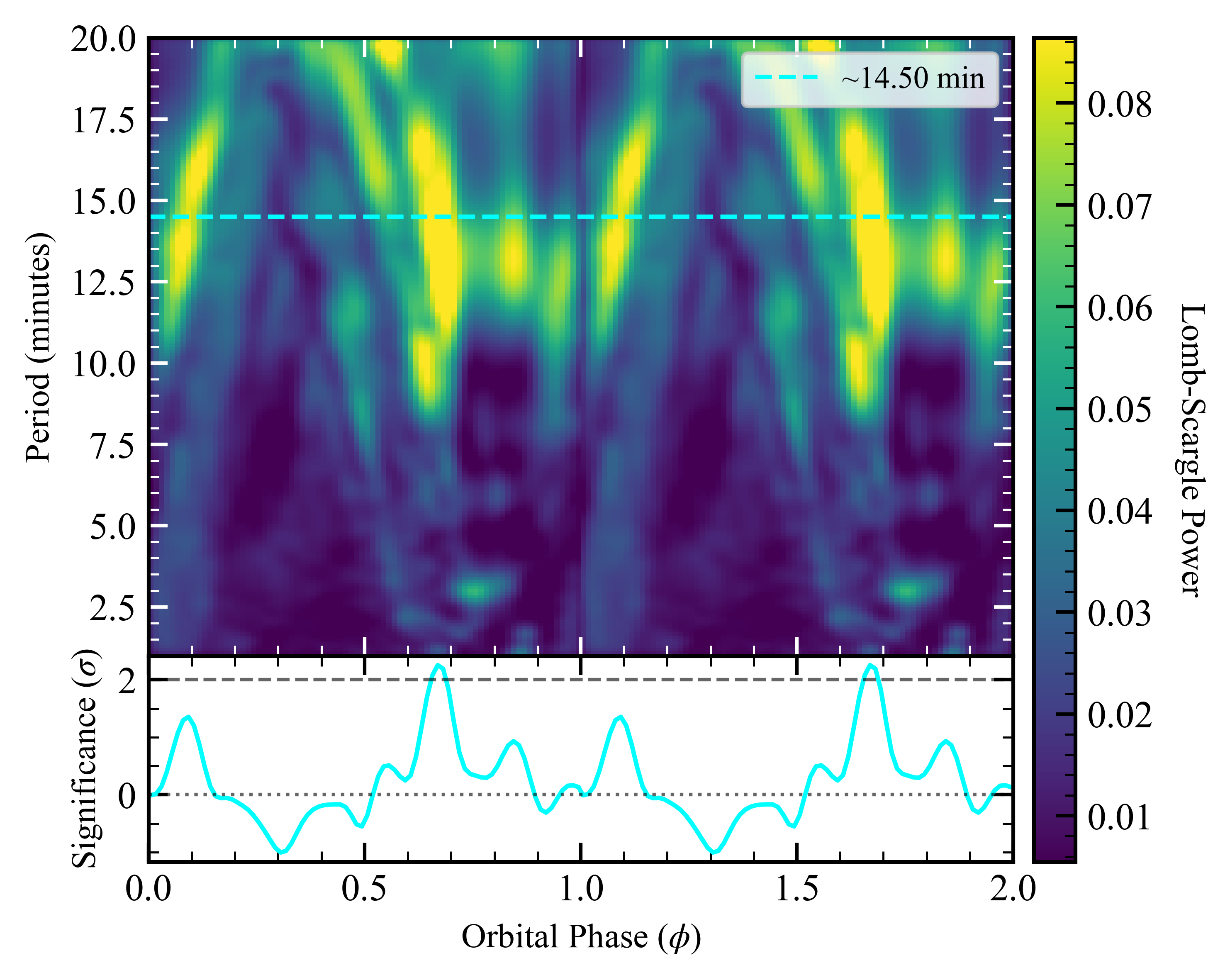}
    \caption{(Top) Lomb–Scargle periodogram of the detrended EPIC pn (0.15-1 keV) light curve. The solid black line shows the power spectrum after removing the long-term trend. (Bottom) The dynamic power spectra of the X-ray light curve showing phase-dependency of the QPO.\label{fig:xray_qpo}}
  \label{fig:xray_qpo}
\end{figure}

\section{Spectral analyses} \label{section:spectra}
\subsection{X-ray}
\subsubsection{Bright phase}
Photons from the bright intervals of each cycle were selected separately and combined, omitting the intervening absorption dips to characterize the time-averaged emission during this phase. The EPIC MOS-1 \& 2 spectra were fitted jointly, first in the energy range 0.15-10 keV, since the emission above 10 keV was negligible and statistically insignificant. The EPIC pn spectrum showed obvious anomalous deviations below 0.25 keV likely from residual effects of pile-up and calibration issues in the softer energies, hence the analysis was restricted to the energy range 0.25-10 keV.

The mean bright-phase spectrum was modeled using a soft blackbody component plus additional components accounting for the unrelated hard, faint emission seen at all phases. The latter was modeled using a multi-temperature plasma model determined in Sect.~\ref{section:faintspec}, the parameters of which were kept frozen during the fitting process. The blackbody temperatures derived from both EPIC MOS and EPIC pn detectors are equivalent, that is, $T_{bb}=35\pm 0.53$~eV and comparable to the values shown by other soft polars, which range from 20 eV to 60  eV \citep{V1309Ori}. The temperature derived here is higher than the temperature estimate from the high-state \textit{EUVE} EUV bright-phase spectral fit of $22.84$~eV \citep{Szkody_EUVE_ASCA_RXTE_1999}. We note, however, that the \textit{EUVE} observation took place during a high state with V~mag of $\sim$15.1 \citep{Szkody_EUVE_ASCA_RXTE_1999} whereas during our observation the V magnitude was exceptionally bright at V$\sim$13.8. The EPIC pn spectrum produces a better fit with reduced chi-square ($\chi_{r}^{2}$) of 1.2, than MOS-1 ($\chi_{r}^{2}$ =1.4). The EPIC MOS-1 spectral fit was improved significantly by introducing a partially covering absorber of column density ($N_{H}$) of $0.15 \times 10^{22}$ cm$^{-2}$ ($\chi_{r}^{2}$ = 0.9) whereas the EPIC pn fit only showed a marginal improvement to a $\chi_{r}^{2}$ =1.1. However, we also note that restricting the lower energy limit of EPIC MOS-1 spectra to 0.25 keV achieves equally good fit without the need for a partially covering absorber. Due to the highly stochastic nature of the emission at soft energies it is hard to discern detector calibration issues at these low energies and intrinsic absorption. Hence we gave preference to the model devoid of the partial absorber, that is, \texttt{Tbabs*(bbody+3apec)}. The MOS-2 spectrum did not produce an acceptable fit in the energy range 0.15-10 keV for any of the models tried and the analysis was restricted to 0.25-10~keV. The spectral-fit parameters are shown in Appendix~\ref{append:spectral_fits} and the best-fit model and data are shown in Fig.~\ref{fig:phase_resolved_spectra}. The unabsorbed, bolometric flux of the soft blackbody, $F_{\rm bb, bol}$ is $6.16 \times 10^{-10}$ erg s$^{-1}$ cm$^{-2}$ and $7.41 \times10^{-10}$ erg s$^{-1}$ cm$^{-2}$ for EPIC MOS-1 and EPIC pn, respectively. The unabsorbed, bolometric flux of the plasma emission components in the bright phase, $F_{\rm plasma, bol}$ is $1.17 \times10^{-12}$ erg s$^{-1}$ cm$^{-2}$ for EPIC MOS-1 and $1.23 \times10^{-12}$ erg s$^{-1}$ cm$^{-2}$ for EPIC pn. This implies a $\frac{F_{\rm bb}}{F_{\rm plasma}}>500$ (526 (EPIC MOS-1) -- 602 (EPIC pn)). However, we note here that the hard emission is interpreted to arise from a different location than the soft blackbody emission seen in the bright phase and is unrelated to the shock (buried or standing) on the WD. Therefore, the soft excess in AR~UMa is essentially infinite as the emission from the primary accreting pole is exclusively in the soft X-rays.
 
The Galactic column density was modeled using \texttt{Tbabs} \citep{Wilms_2000} with the abundance table set to \texttt{vern} \citep{vern}. When left as a free parameter, the neutral hydrogen column density  ($N_{\rm H, tbabs}$) is estimated to be $1 \times 10^{19}$ cm$^{-2}$. While this is an order of magnitude higher than the estimate from the low-state \textit{HST} STIS spectrum \citep{STIS_ARUMa}, it is consistent within errors with the estimate from the high-state \textit{EUVE} spectrum \citep{Szkody_EUVE_ASCA_RXTE_1999}. We note, however, that due to the insensitivity of the model to this negligible absorption, fixing $N_H$ to zero or to the \textit{HST} STIS value yields statistically equivalent fits without significantly altering the remaining model parameters.

We also investigated the high-resolution RGS spectra for emission and absorption features in the bright phase. We utilized the  RGS-1 \& 2 combined spectra and restricted our analysis to the first order. The continuum emission in RGS is described the same way as for the EPIC instruments with an addition of an emission line at 0.404 keV, with an equivalent width of 4 eV, identified with Fe \textsc{xix}. The best-fit spectral parameters are shown in Appendix~\ref{append:spectral_fits} and the best-fit model and data are shown in Fig.~\ref{fig:phase_resolved_spectra}. We also investigated cycle-to-cycle variation in the blackbody temperature using the RGS bright-phase spectra and is described in Appendix \ref{append:cycle_bb_var}.

\subsubsection{Faint phase}\label{section:faintspec}
\begin{figure}
    \centering
        \centering
            \includegraphics[width=0.41\textwidth]{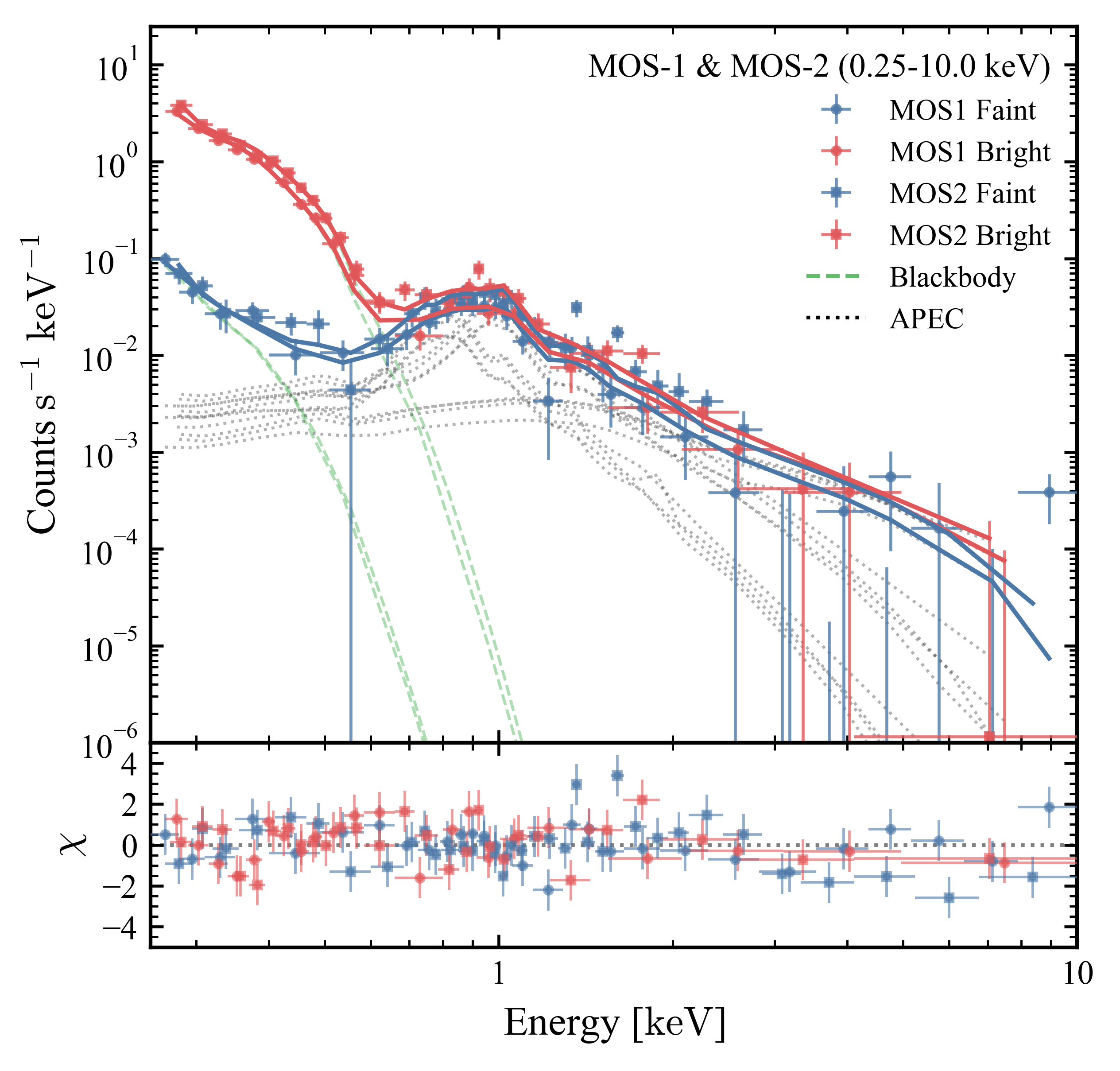}
    \vspace{-0.15cm}
    \includegraphics[width=0.41\textwidth]{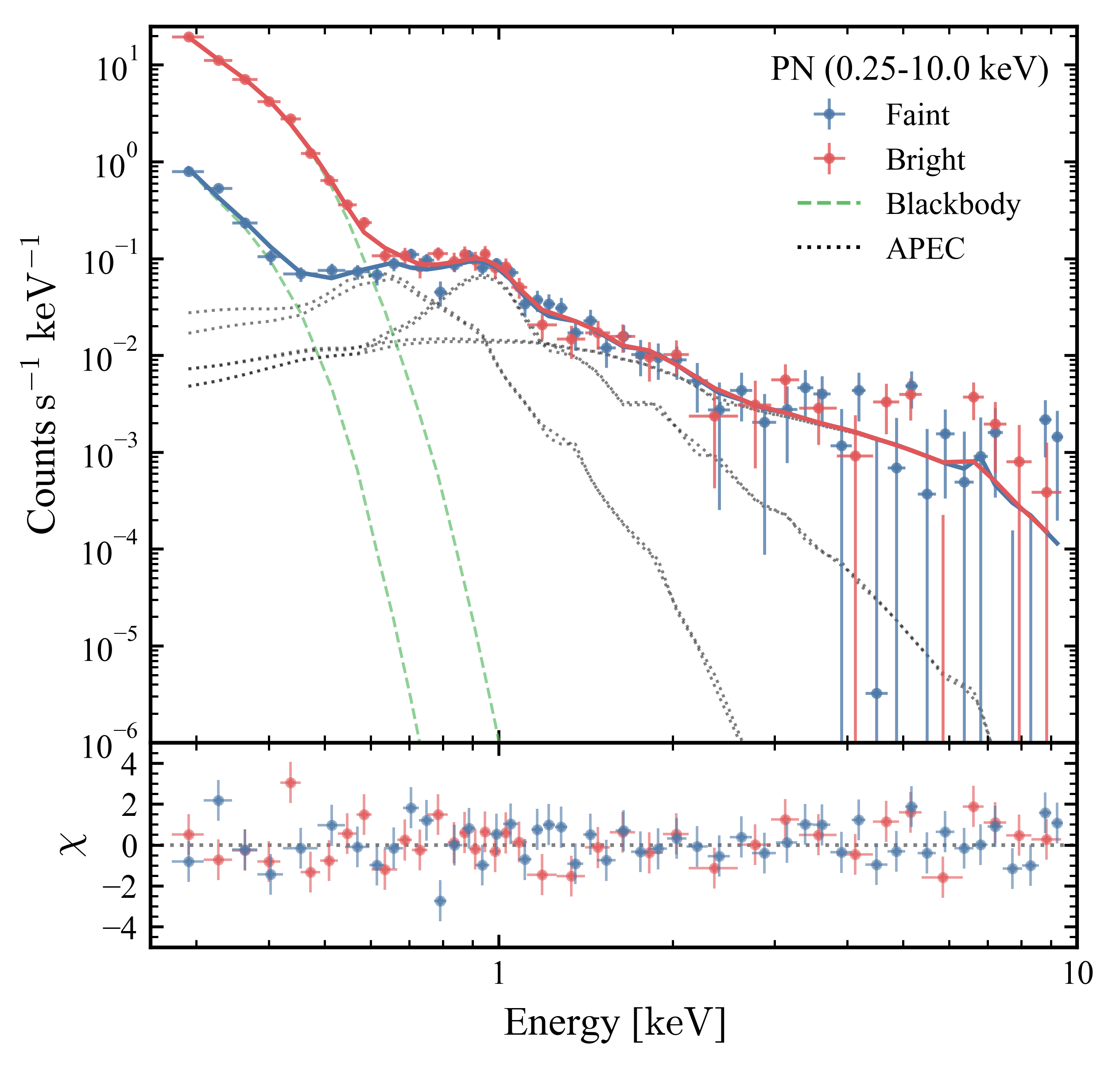}
        \vspace{-0.15cm}
        \centering
    \includegraphics[width=0.41\textwidth]{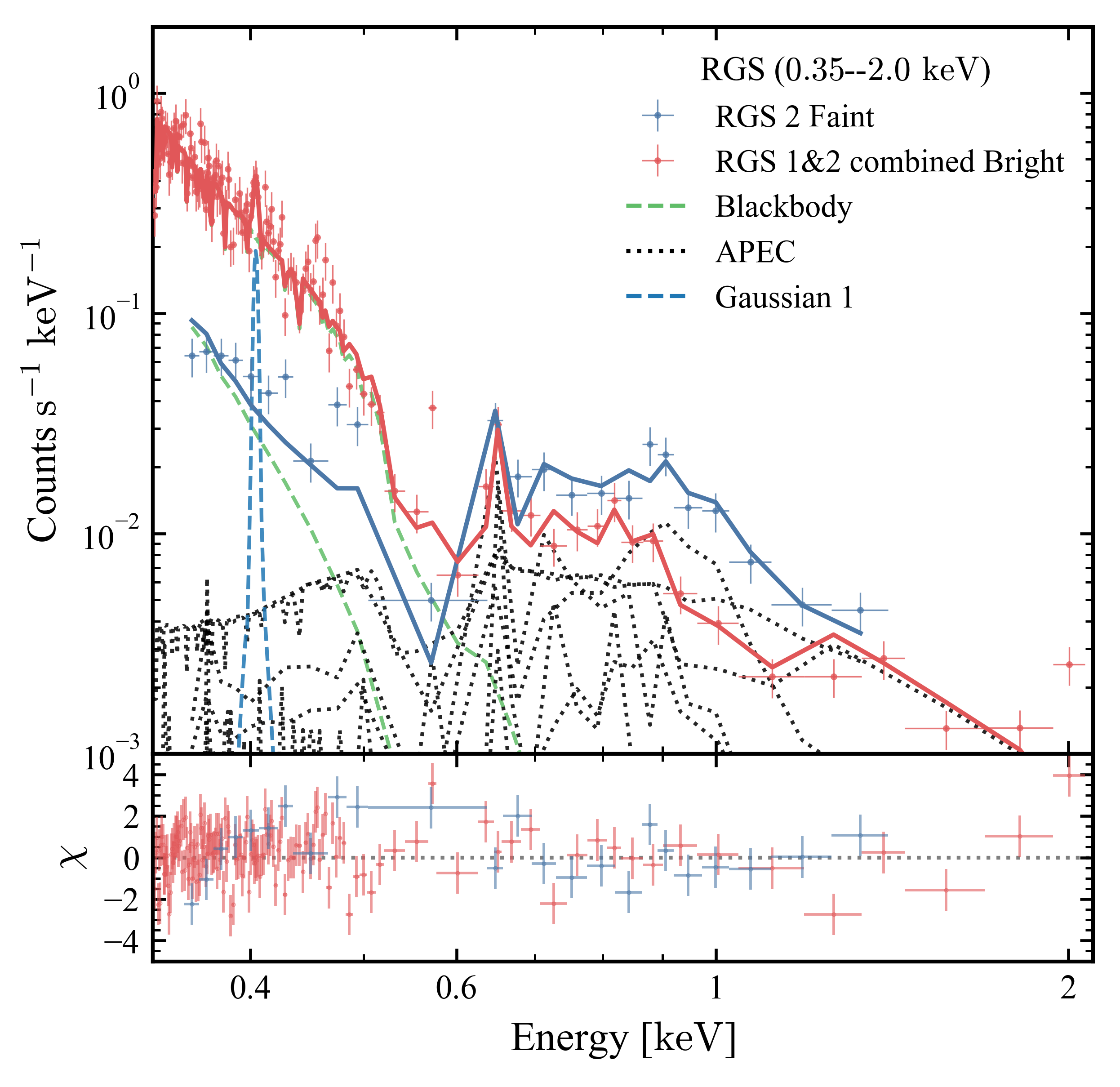}
    \caption{Phase-resolved spectral analysis of the faint and bright phases for (top) EPIC MOS-1 + MOS-2 (0.25-10~keV), (middle) EPIC pn (0.25-10~keV), and (bottom) RGS (0.35-2.0~keV). For RGS bright-phase spectroscopy RGS-1 \& 2 data were combined whereas for the faint phase only RGS-2 data were used. The faint-phase data are shown in blue and the bright-phase data are shown in red. The solid lines show the respective best-fit models whereas the dotted lines show the individual components. }
    \label{fig:phase_resolved_spectra}
\end{figure}

The faint-phase spectroscopy was carried out by combining the photons from faint intervals of each cycle. The mean faint-phase spectrum shown in Fig. \ref{fig:phase_resolved_spectra} is much fainter and flatter than the bright-phase spectrum and comprises the hard tail (above $\sim$1 keV) of the latter entirely. The faint-phase spectrum cannot be modeled either by a single blackbody or a single plasma emission component or a combination of both. Instead the shape of the faint emission is suggestive of multi-temperature optically thin emission, which we modeled using two plasma (\texttt{apec}) components (we also added a phenomenological \texttt{apec} component frozen at 10 keV to account for the low-count hard energy tail). The residuals from this multi-temperature fit are confined to energies below 0.3 keV and can be fitted by a blackbody of slightly lower temperature than that of the bright-phase spectrum, albeit a plasma component of same temperature can also be used to fit this soft component. A weaker, softer blackbody in the faint phase as compared to the bright phase is consistent with the EUV spectral fits from \cite{Szkody_EUVE_ASCA_RXTE_1999}, however similar to the bright-phase blackbody temperature, the faint phase estimate, that is, of $\sim$25-27 eV is higher than what is estimated from EUV spectrum ($18.5$ eV). The best fit for this phase has a satisfactory $\chi_{r}^{2}$ of 0.65 for EPIC MOS-1 without the addition of a partial absorber ($\chi_{r}^{2}$=0.63), for EPIC pn these values are 0.97 and 0.94, respectively. The unabsorbed, bolometric flux of this plasma component is $F_{\rm plasma, bol}$ = $9.3 \times10^{-13}$ erg s$^{-1}$ cm$^{-2}$ from EPIC MOS-1 best fit and is comparable within errors to the estimates from EPIC pn of $1.17 \times10^{-12}$ erg s$^{-1}$ cm$^{-2}$ and is equal to the plasma flux measured in the bright phase. The best-fit spectral parameters are shown in Appendix~\ref{append:spectral_fits}.  

The faint-phase RGS-1 spectrum shows a spurious excess centered at 0.387 keV which is absent in RGS-2 (see Appendix \ref{rgs_anomaly_append}), therefore only RGS-2 spectrum was considered for analysis. Although the general spectral shape is consistent with that of EPIC instruments, RGS-2 faint-phase spectrum shows excesses that are not seen in the EPIC data (see Fig. \ref{fig:phase_resolved_spectra}). However, due to the low counting statistics during the faint-phase interval, we did not fit additional spectral components to avoid over-interpretation. The best-fit faint-phase model parameters for RGS-2 are given in Appendix~\ref{append:spectral_fits}.

\subsubsection{Stream dip}\label{subsubsection:stream-dip}
\begin{figure}[ht!] 
    \centering
        \includegraphics[width=\columnwidth]{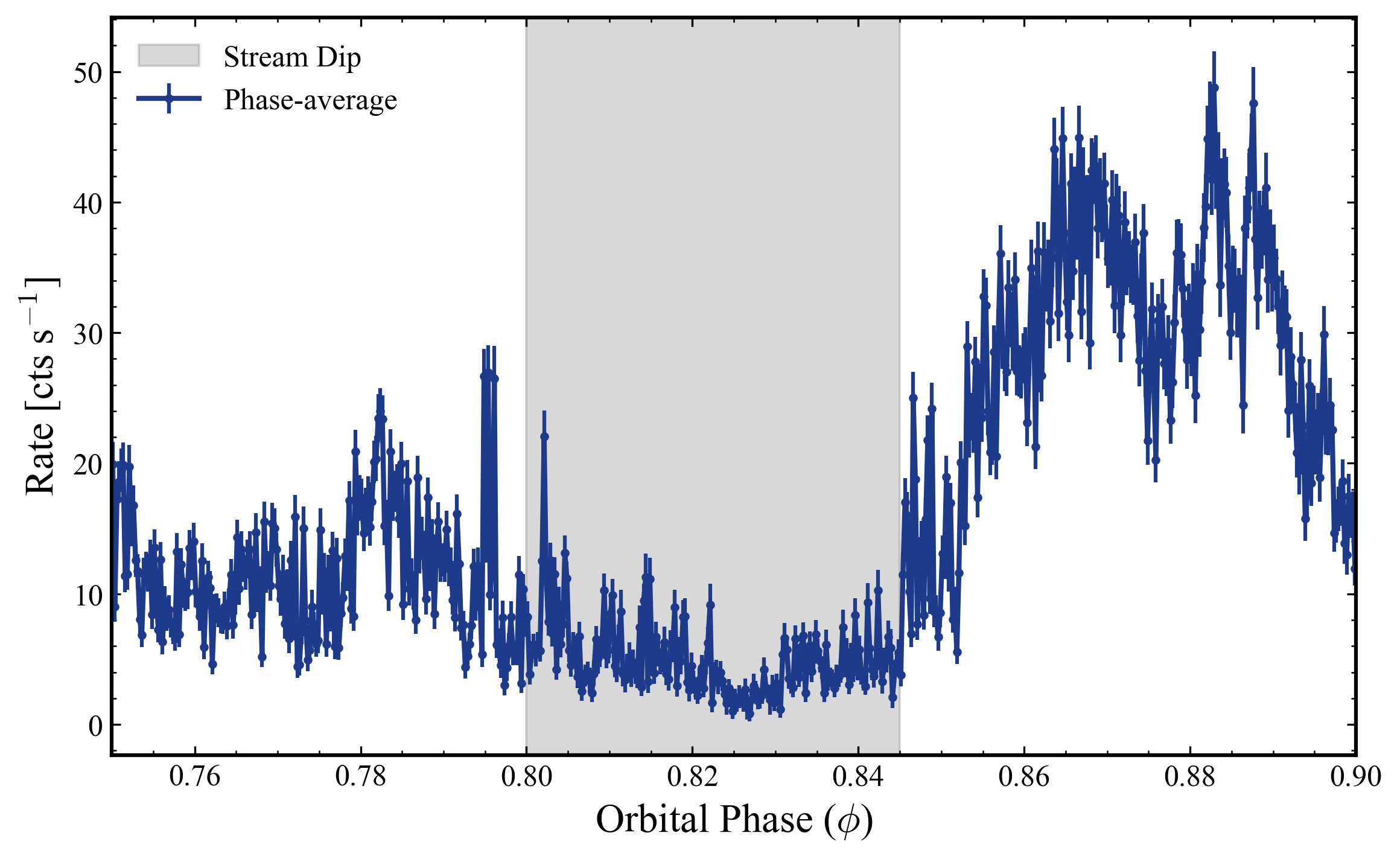}
    \vspace{-0.15cm} 
        \includegraphics[width=\columnwidth]{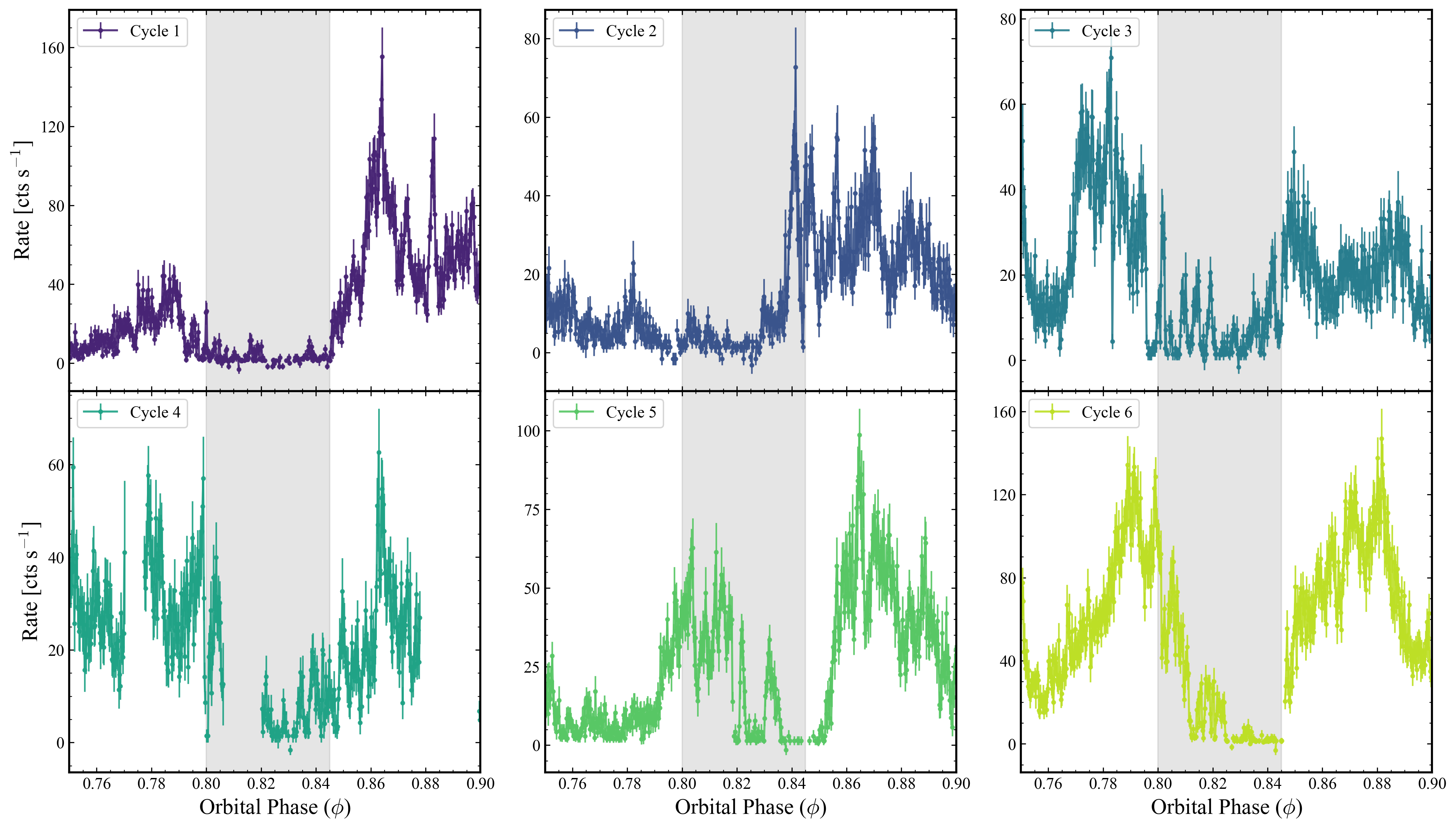}
        \caption{(Top) \textit{XMM-Newton} EPIC pn (0.2-1 keV) phase-averaged light curve showing the mean stream-dip interval $\phi \simeq 0.80-0.845$ (shaded in gray). (Bottom) The panels illustrate the meandering of stream dip each cycle, shaded in gray is the phase-averaged position of the stream dip.}
    \label{fig:stream-dips}
\end{figure}

\begin{figure}[ht!] 
    \centering
            \includegraphics[width=0.9\columnwidth]{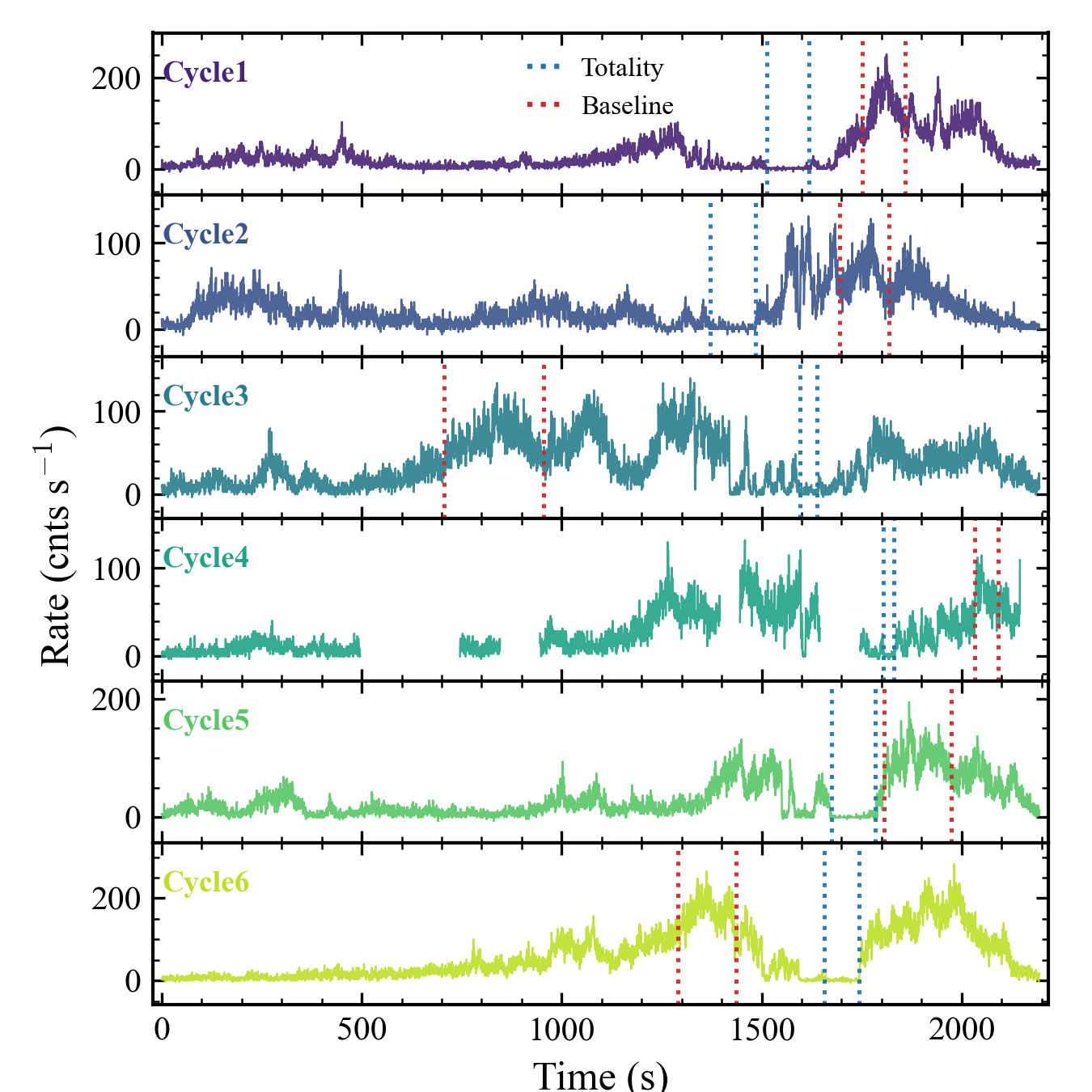}
             \caption{High-resolution (1~s) EPIC pn (0.15--1~keV) light curves illustrating the bright intervals across the six observed orbital cycles of AR~UMa. The blue dotted regions denote the least active period (the total stream-dip phase), while the regions denoted by the red dotted lines encompass the baseline of each cycle considered to be least obscured.}
              \label{fig:bright_phase_cycles}
    \end{figure}

\begin{figure}
    \centering
   \includegraphics[width=1\columnwidth]{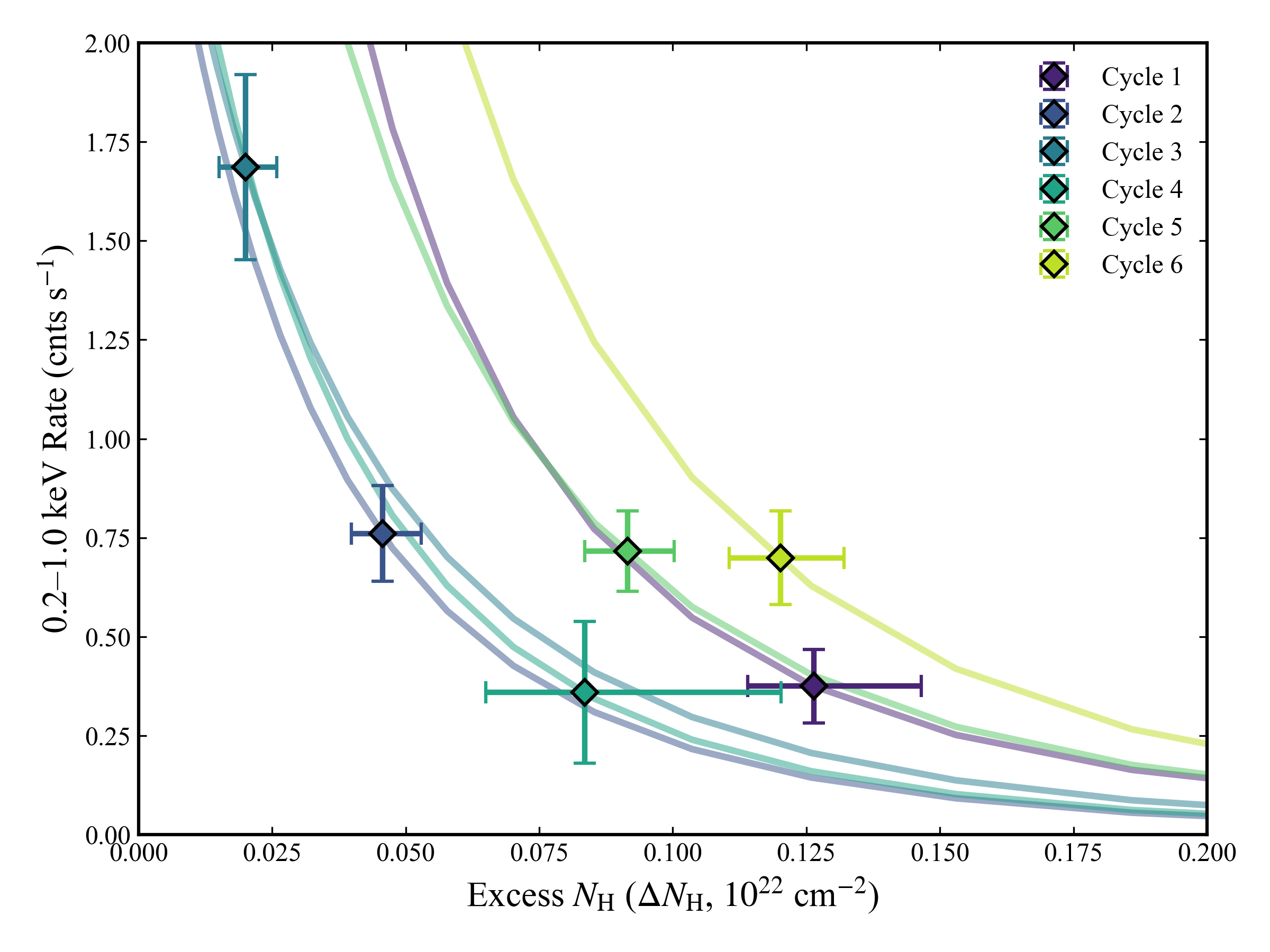}
    \caption{Variable density profile of the obscuring stream, suggesting the inhomogeneous nature of the accretion flow. The $N_{H}$ is interpolated from best fit of the baseline interval of respective cycle using count-rate of the coincident stream-dip totality, the intervals are marked in Fig.~\ref{fig:bright_phase_cycles}.}
    \label{fig:streamdip_nh}
\end{figure}

The stream-dip extracts from the six full orbital cycles covered by \textit{XMM-Newton} are shown in Fig.~\ref{fig:stream-dips}. The phase-averaged stream dip, shown in the first panel, shows a very broad V-shaped dip which extends from $\phi \approx 0.80-0.845$ and is marked in gray. In each of the six stream dips shown in the bottom panel the phase-averaged position of the stream dip is marked. Within each of the six there is substantial substructure, usually close to the edge of the dip, characterized by variability on time scales ranging from 2-70~s. The count rate within the dip ranges from almost zero to values comparable with average of the bright phase. In some cycles the dips are almost total in a narrow phase range, $\Delta \phi \approx 0.02$, but the centroid of this ``totality'' meanders from cycle to cycle. In order to estimate the density fluctuations in the stream we derived cycle-specific column density. We used the best-fit model \texttt{Tbabs*(bbody+apec)} in the energy range 0.2-1 keV, for the  brightest interval, henceforth ``baseline,'' of a given cycle and increased the absorption until the model count rate is equal to that of the observed count rate in the respective dip minimum. In Fig.~\ref{fig:bright_phase_cycles} we demarcate the baseline interval and totality intervals used for the analysis. The column density thus estimated for each cycle is shown in Fig.~\ref{fig:streamdip_nh} and tabulated in Tab.~\ref{table:stream_nH} and is direct evidence of an inhomogeneous flow. We note here that the column densities determined thus are only lower limits considering there exists a persistent emission which also comprises a soft component. 

\begin{table}
\caption{Variable density profile of the accretion stream derived for six orbital cycles of AR~UMa.}
\label{table:stream_nH}
\centering
\footnotesize
\setlength{\tabcolsep}{3pt}
\begin{tabular}{cccc}
\hline\hline
\noalign{\smallskip}
Cycle & Baseline Count Rate & Dip-totality Count Rate & Excess $\Delta N_{\rm H}$ \\
      & ($\rm cts\ s^{-1}$) & ($\rm cts\ s^{-1}$) & ($10^{21}\ \rm cm^{-2}$) \\
\noalign{\smallskip}
\hline
\noalign{\smallskip}
1 & $77.93 \pm 1.02$ & $0.38 \pm 0.09$ & $1.26_{-1.1}^{+1.5}$ \\
2 & $31.04 \pm 0.61$ & $0.76 \pm 0.12$ & $0.46_{-0.4}^{+0.5}$ \\
3 & $34.65 \pm 0.45$ & $1.69 \pm 0.23$ & $0.20_{-0.1}^{+0.2}$ \\
4 & $36.29 \pm 0.94$ & $0.36 \pm 0.18$ & $0.83_{-0.6}^{+1.2}$ \\
5 & $54.37 \pm 0.68$ & $0.72 \pm 0.10$ & $0.91_{-0.8}^{+1.0}$ \\
6 & $95.20 \pm 0.96$ & $0.70 \pm 0.12$ & $1.20_{-0.1}^{+0.1}$ \\
\noalign{\smallskip}
\hline
\end{tabular}
\tablefoot{
All count rates are extracted from the EPIC pn light curve in the soft $0.2- 1.0$ keV range. 
The total column densities ($N_{\rm H}$) are computed via grid-based count-rate interpolation using the frozen best-fit parameters derived from each cycle's respective unabsorbed bright-phase interval (baseline), incorporating a fixed Galactic extinction baseline $N_{\rm H, Gal} = 1\times 10^{19}\ \rm cm^{-2}$.
}
\end{table}

\subsection{UV}
\begin{figure}[ht!]
    \centering
\includegraphics[width=\columnwidth]{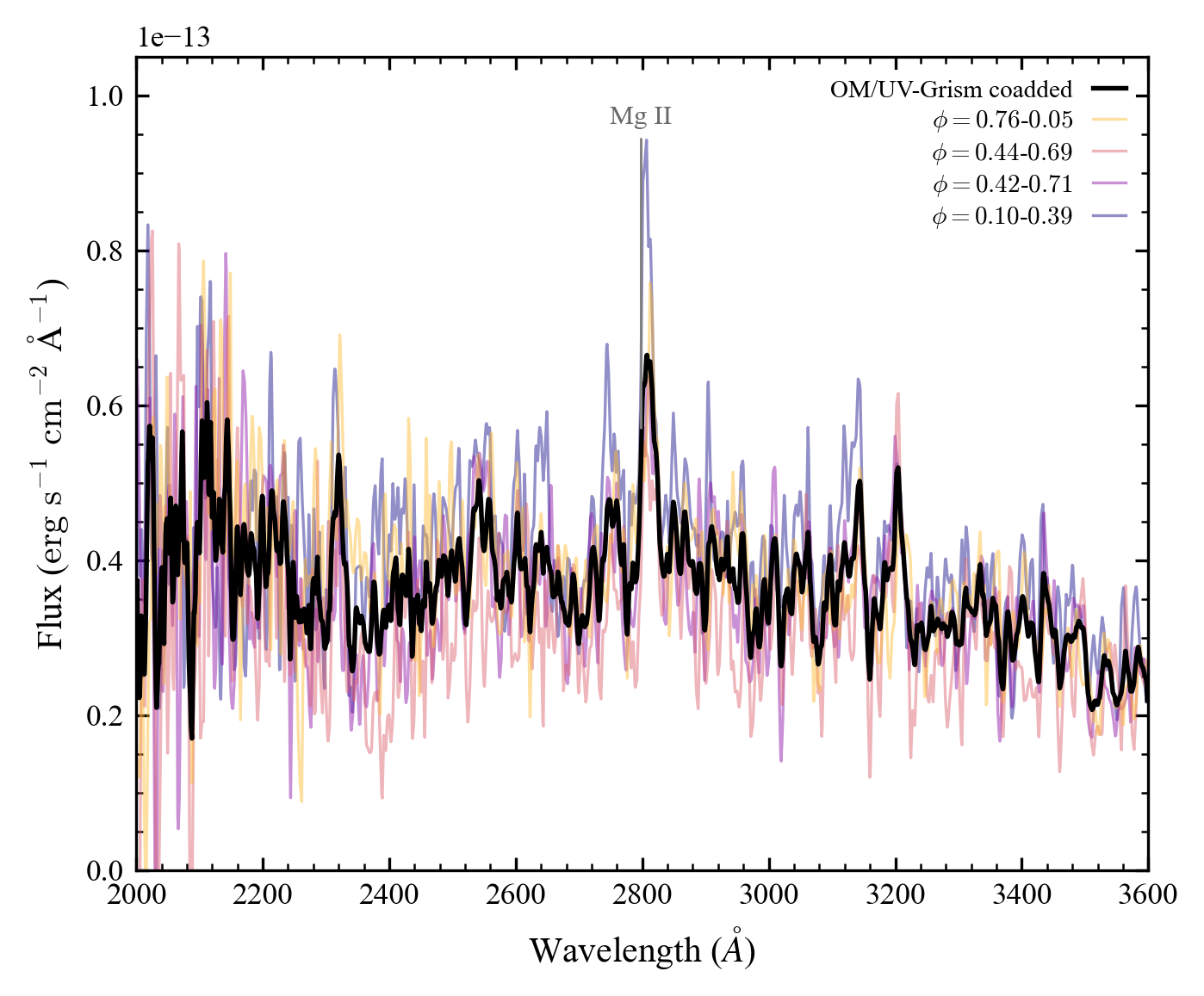}
 \caption{\textit{XMM-Newton} OM-UV Grism (2000-3600\ $\AA$) coadded spectra (black) overlaid on individual sub-exposures identified in the legend.}
 \label{fig:om-grism_coadd}
 
\includegraphics[width=\columnwidth]{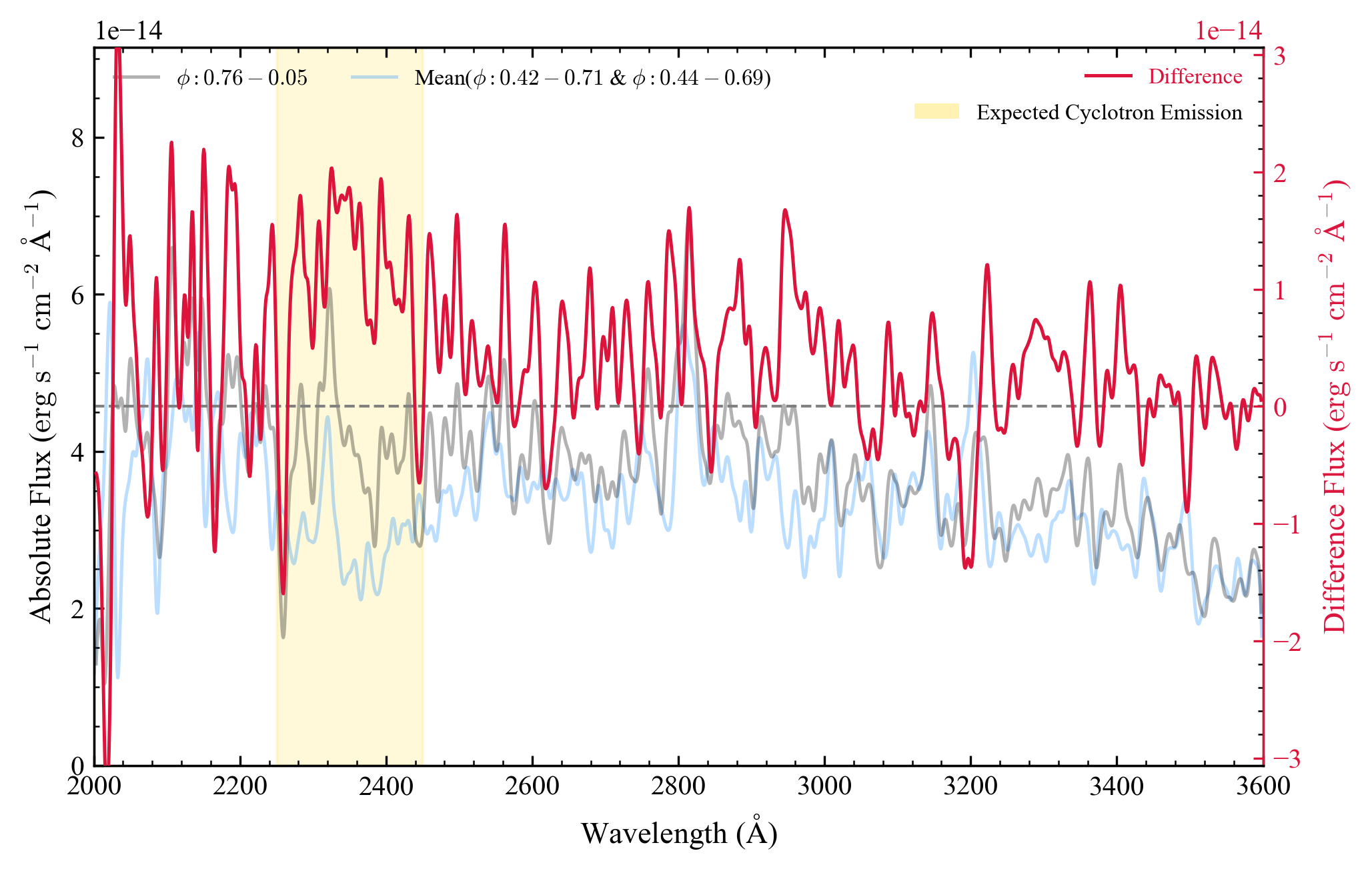}
    \caption{\textit{XMM-Newton} OM UV-Grism difference spectrum showing an excess between 2250-2450 $\AA$, identified with cyclotron emission from the third harmonic of the weak southern (160~MG) pole \citep{STIS_ARUMa}. The panels show difference spectra between spectrum from the phase interval $\phi$ (0.76-0.05) and mean of the sum of spectra from $\phi$ = (0.42-0.71) and (0.44-0.69). The difference spectrum is shown in red and the phase-resolved spectra are shown in blue and black translucent lines.}
    \label{fig:diff}
\end{figure}

The \textit{XMM-Newton} OM UV-Grism (1800-3600 $\AA$) spectra obtained as four separate exposures were coadded to produce the average spectrum. Since the sensitivity of the OM UV-Grism drops drastically below 2000$\AA$ the spectra were analyzed only in the range 2000-3600 $\AA$. Figure~\ref{fig:om-grism_coadd} shows the coadded spectrum overlaid on the four sub-exposures. The individual exposures were phased and investigated separately to identify phase-dependent variations. The four exposures belong to phases $\phi$ = 0.76-0.05, 0.44-0.69, 0.42-0.71, 0.10-0.39. Apart from the marked emission lines, there are no obvious signatures especially that of cyclotron harmonics in either the average spectrum or the individual exposures, even after continuum subtraction. However, difference spectra obtained by subtracting the spectra from phase, $\phi$ $\sim$ 0.4-0.7 from that of the spectrum from phase $\phi$ $\sim$ 0.76-0.05 (when the southern pole is the closest) show excess emission between $\lambda$= 2250--2450 $\AA$ which is shown in Fig.~\ref{fig:diff}. An excess at 2230 $\AA$ was predicted from cyclotron modeling by \cite{STIS_ARUMa} arising from cyclotron emission from the second and third harmonic from the stronger (230~MG) and weaker (160 MG) poles, respectively. Given this excess is only seen in the difference spectra of the X-ray bright phase, that is, when the southern pole is the closest, it is interpreted to be cyclotron emission from the southern pole. We also compare the high-state UV-Grism spectra with that of the low-state UV and optical, and high-state EUV and optical spectra to characterize the spectral energy distribution of the system in Appendix~\ref{append:SED}.

\section{Results and discussion} \label{results_discussion}
We report the first high-state \textit{XMM-Newton} observation of the highest field polar, AR~UMa, covering $\approx 6.5$ orbits of its 1.93 h cycle.  We have carried out multiwavelength timing analysis, in the soft and hard X-rays using \textit{XMM-Newton} EPIC pn, in the UV using \textit{XMM-Newton} OM-UVW1 and in the optical using the 70-cm telescope at the AIP. We also characterize the spectral features of the system in the X-ray and UV.

The X-ray light curve of the source is heavily modulated into a bright phase, lasting 35\% of the orbital cycle, and a faint phase. The energy-resolved light curves show that the bright-phase emission is almost entirely soft (below 1keV) and flaring, with no correlated high-energy photons that could be attributed to a standing shock. The soft, bright-phase emission is interrupted by strong absorption events, the more stable and pronounced of which is interpreted to be the line-of-sight obscuration of the accretion area by puffed-up matter in the stagnation region. The faint, hard X-ray emission (above 1 keV) is not modulated over the orbital cycle, and is unlike the bremsstrahlung emission seen in canonical polars which is expected to have temperatures of several tens of keV \citep{1985Lamb}. Phase-resolved X-ray spectroscopy reveals a soft blackbody (35 $\pm$ 0.53 eV) in the bright phase and a persistent faint plasma emission ($\sim$ 1 keV) seen throughout the orbital cycle. The stream dip shows dramatic variations cycle to cycle both in its phasing and activity levels indicating a porous stream showing density fluctuations consistent with the picture of a clumpy flow. Using the depleted soft counts seen in this interval we estimate cycle-to-cycle column density of the stream as direct measure of inhomogeneity in the flow.

The optical and UV light is dominated by the stream with possible contributions from the irradiated surface of the secondary and is antiphased to the X-ray emission. Both light curves show stochastic flaring and signatures of QPOs. Power spectral analysis reveals a $\sim$4-5 min QPO in UV and optical which is pronounced in the orbital-phase interval $\phi\sim0-0.3$ and $\phi\sim0-0.5$ respectively, when the primary, soft-X-ray emitting pole is behind the limb of the WD. These timescales are too long compared to the dynamical timescales of individual filaments which are expected to be seconds long and might result from superposition of excitations driven by individual blobs. Considering the UV and optical light is predominantly from the stream a more plausible origin could be the oscillation of the accretion column formed in the stagnation region. The 12.6 min long QPO identified in the optical is likely a harmonic of the 5.1 min long signal. The high-state UV spectrum reveals cyclotron emission arising from the third harmonic of the weaker southern pole of field strength 160 MG. In the following sections we determine the energy budget of the source and the instantaneous mass transfer rate and discuss the potential origin of the persistent faint hard emission.

\subsection{The accretion rate and energy budget of AR~UMa}

To quantify the energy budget of AR~UMa we calculate the unabsorbed soft and hard X-ray luminosities ($L_{\mathrm{bb}}$ and $L_{\mathrm{plasma}}$) for a distance of $d = 98.62\pm0.52$ pc, averaged across instruments. The soft X-ray luminosity, $L_{\mathrm{bb}}$, is calculated for two different geometric assumptions: a standard half-hemispheric emission ($L_{\mathrm{bb}}$ = $2\pi \mathrm{F}_\mathrm{{bb}} \mathrm{d}^{2}$) and a geometrically corrected emission accounting for the system inclination $i = 70^\circ$ and the spot colatitude, $\beta = 142^\circ$, that is, $L_{\mathrm{bb}}$ = $\frac{1}{\mathrm{cos}(\beta-i)}\pi \mathrm{F}_{\mathrm{bb}} d^{2}$ $\approx$ $3.2\pi \mathrm{F}_{\mathrm{bb}} d^{2}$, giving the lower and upper bounds of estimates, respectively. The $L_{\mathrm{plasma}}$ is calculated for an isotropic emission as $4 \pi \mathrm{F}_{\mathrm{plasma}} \mathrm{d}^2$. The total X-ray luminosity ($L_{\mathrm{X}}$) is heavily dominated by the soft blackbody component with $L_{\mathrm{bb}}$ = 3.95-6.39 $\times$ 10$^{32}$ erg s$^{-1}$ and $L_{\mathrm{plasma}}$ = 1.4 $\times$ 10$^{30}$ erg s$^{-1}$, resulting in a soft-to-hard ratio of 282-456. As noted previously, the hard X-ray emission seen in AR UMa is uncorrelated with the bright-phase soft emission and we assume a completely buried shock, thus the ratio derived is suggestive only in terms of the total energy budget of the system and does not indicate degree of reprocessing as it would apply for a standing shock model. Assuming that the total X-ray luminosity traces the accretion luminosity (as cyclotron luminosity, $L_{cyc}$ in the high state is negligible) ($L_{\mathrm{X}} \approx G M_{\mathrm{wd}} \dot{M} / R_{\mathrm{wd}}$), we derived the mass transfer rate ($\dot{M}$) adopting a WD mass of $M_{\mathrm{wd}} = 1.125 \ M_\odot$ \citep{ARUMa_TESS} and a corresponding radius of $R_{\mathrm{wd}} = 4.48 \times 10^8$ cm \citep{Nauenberg}. We find a mass accretion rate of $\dot{M} \approx (1.19 - 1.92) \times 10^{15} \ \mathrm{g\,s^{-1}}$, corresponding to $(1.18 - 3.05) \times 10^{-11} \ M_\odot\,\mathrm{yr}^{-1}$.    

\subsection{Nature of accretion flow in AR~UMa} \label{accretion_flow}
AR~UMa shows an extreme case of soft blobby accretion with the active pole emitting exclusively in the soft X-rays. This pronounced soft X-ray emission is incompatible with the canonical picture of accretion in magnetic CVs, that is, of a standing shock above the WD. Shock-heating is unavoidable for a supersonic flow, however the hard X-rays can effectively be culled if the shock is buried in thick photospheric layers of the WD. The soft X-ray dominant MCVs are thus understood to be accreting matter via dense, diamagnetic blobs or rather, field-aligned filaments that bury their shock. It is envisaged that the transferred matter will enter the magnetosphere in the form of clumps through the development of various interchange instabilities such as the Rayleigh-Taylor (RT) instability as first discussed in the context of spherically symmetric accretion onto neutron stars by \cite{Blobs_NeutronStars}. 

The accretion flow that leaves the L1 does so at ballistic speeds and is essentially diamagnetic since it is able to avoid the field completely by developing surface currents. The trajectory of the inflow remains ballistic until the ram pressure, $\rho_{s}v_{\rm ff}(r)^{2}$, of the inflow equalizes with the magnetic pressure, $\frac{B^{2}(r)}{8\pi}$. We can estimate the stagnation point (the stream-magnetospheric radius at which the inflow is influenced by the magnetic field) as \citep{1985Lamb}
 \begin{equation}
     R_{\mu}^{st} = 1.75 \times10^{10}\eta^{4/11}(B^{p}_{7})^{4/11} \dot{M}^{-2/11}P_{\rm orb,5}^{4/11} {M}_{1}^{-1/11}R_{\rm wd,9}^{12/11}\ \mathrm{cm},
 \end{equation}
where $\eta$ is a dimensionless parameter of order unity, $B^{\rm p}_{\rm 7}$ is the effective polar magnetic field strength in units of $10^7 \ \mathrm{G}$, $\dot{M}$ is the mass transfer rate in units of $10^{16} \ \mathrm{g \ s^{-1}}$, $P_{\rm orb,5}$ is the orbital period in units of $5 \ \mathrm{h}$, ${M}_{1}$ is the WD mass in units of $M_{\odot}$, and $R_{\rm wd,9}$ is the WD radius in units of $10^9 \ \mathrm{cm}$.

Substituting the appropriate values for AR UMa we obtain $R_{\mu}^{st}$ = 2.15-2.35 $\times$ 10$^{10}$ cm. For a binary separation $a$ = 5.92 $\times$ 10$^{10}$ cm and L1 distance of 4.13 $\times$ 10$^{10}$ cm \citep{Warner}, this implies a magnetically channeled flow 50-55\% of the length of the accretion flow. At the stagnation point we expect a shock to form (see Sect.~\ref{subsection:threadingshock}), which increases the density of the inflow by a factor of 4. As the post-shock subsonic flow cools from approximately 4 $\times$ 10$^{7}$ K to 2 $\times$ 10$^{4}$ K, the density increases again by almost 2000 times \citep{1986HKL} at the magnetospheric boundary giving rise to Rayleigh-Taylor instability. The density of the cooled post-shock flow ($\rho_{sh}$) at the magnetospheric boundary can be estimated as \citep[see also][]{1986HKL}  

\begin{equation}
    \rho_{\rm sh} = 14 \times 10^{-8} \eta^{-2} \dot{M}_{16} P_{\text{orb},5}^{-2} {M}_1^{-1/2} \left(R_{\mu,10}^{\text{st}}\right)^{1/2}\  \mathrm{g\ cm}^{-3}.\
\end{equation}
For AR~UMa $\rho_{\rm sh}$ is estimated to be (1.61-2.5) $\times$ 10$^{-7}$ g cm$^{-3}$.

Due to the ensuing exchange instabilities we expect the dense flow to break up into blobs or clumps as they enter the magnetosphere. Further, the filaments that are perpendicular to the field are compressed as they move along the radial distance, $r$, toward the WD, as $(R_{\rm coup}/R_{\rm wd})^{2.5}$. In AR~UMa, the compression factor is (1.6-2) $\times$ 10$^{4}$ which implies a filament density of (3.2-4) $\times$ 10$^{-3}$ g cm$^{-3}$ at the WD surface. This corresponds to a specific mass accretion rate, $\dot{m}_{\rm blob}$ of (2.62-3.25) $\times$ 10$^{6}$ g cm$^{-2}$ s$^{-1}$ which is much higher than the critical specific mass accretion rate of 10$^{2}$ g cm$^{-2}$ s$^{-1}$, according to \cite{Kupiers_Pringle}, above which we expect the shock to be buried within the optically thick photospheric layers of the WD. Considering we detect no hard X-ray photons that could arise in a standing shock it could very well be that the accretion in AR~UMa is entirely in the form of blobs. 

\subsection{On the origin of persistent faint emission} \label{subsection:threadingshock}
During all orbital phases of the source there exists a weak plasma emission of temperature $\sim$ 1 keV that is uncorrelated with the soft emission. The spectral and temporal behavior of this emission is complex and atypical of hard X-ray emission expected from bremsstrahlung radiation. We interpret this hard X-ray emission to be resulting from a stream-magnetospheric shock. As mentioned earlier, the matter that leaves L1 does so in a ballistic trajectory until the magnetic pressure becomes comparable to the ram pressure of the inflow at the stagnation region which we equate with $R_{\mu}^{\rm st}$. At this radius a shock is expected to form, the temperature of which can be estimated as $T_{\rm sh}^{\rm st}$ = 3.7 $\times$ 10$^{7}$ $\frac{{M}_{1}}{R_{\mu,10}^{st}}$ K \citep{1986HKL}, which corresponds to $\approx$ 1.3 keV for AR UMa. At these temperatures the cooling is likely dominated by bremsstrahlung. The cooling length of this flow can be calculated as $D_{\rm s} = 2 \times10^{9}\eta^{2}\dot{M}_{16}^{-1}P_{\rm orb,5}^{2}{M}_{1}^{3/2}(R_{\mu,10}^{\rm st})^{-3/2}$ cm \citep{1986HKL}. Substituting the values for AR UMa the cooling length is estimated to be (6-8)$\times$ 10$^{8}$ cm. We expect anywhere from 3\% to 10\% of the accretion luminosity to be dissipated as a result of this shock. In polars with standing shock however, we expect this emission to be outshone by the shock-heating in the column. It becomes visible here so clearly due to the preferential geometry of the system and the extreme soft X-ray emission from the main accretion region. The emission of this kind has only been discussed in case of V1309 Ori which shows a purely blobby accretion, and as such the fact that we are able to detect this emission in AR~UMa strengthens the picture of buried shocks. 

Besides the aforementioned hard X-ray component, the faint phase also shows a soft blackbody component which contributes the largest fraction of the bolometric flux in this phase. The blackbody temperature of this component (26 $^{+2.0}_{-1.6}$ eV) is much softer than the blackbody temperature seen in the bright phase (35 $^{+0.5}_{-0.6}$ eV) but the emitting area is much smaller, that is, $A_{bb}$ $\sim$ 7 $\times$ 10$^{13}$ cm$^{2}$ in the faint phase and $\sim$ 3-5 $\times$ 10$^{14}$ cm$^{2}$ in the bright phase. These properties are identical to those observed in the EUV \citep{Szkody_EUVE_ASCA_RXTE_1999}. We observe an orbital modulation in this component with the flux increasing toward the bright phase. This prompts us to favor the interpretation that this emission arises from a cooler, extended accretion region on the southern pole which is visible even when the hottest regions are behind the limb of the star. In a clumpy flow, blobs that originate in the stream-magnetospheric boundary have different masses and sizes. The smaller mass clumps diffuse into the field lines almost immediately. Conversely, the larger mass clumps continue to follow dynamical, ballistic trajectories deeper into the magnetosphere. As these larger blobs travel, they are stripped by the Kelvin-Helmholtz (KH) instability into finer drops or mists that easily couple to the field lines. Because the densities of the blobs vary, they couple to the magnetic field over a range of depths. As a result, the footprint of the flow is not a symmetrical, circular spot as would be expected from a smooth, collimated flow, but is instead extended over a range of azimuth \citep{Kupiers_Pringle, Wikramsighe_arc}. Another plausible origin for this soft emission is accretion onto the strongest, northern pole. However, we disfavor this picture for the following reason: since there are no hard photons arising from a standing shock at the northern pole either, accretion onto this pole would also have to be heavily clumped (blobby). However, unlike the flaring and flickering characteristic of an inhomogeneous flow (for example as seen in the bright phase, Fig.~\ref{fig:bright_phase_cycles}), the soft emission in the faint phase is remarkably stable and devoid of any dramatic variations.

\subsection{Characterizing the stream dip} \label{stream_geometry}
A pronounced absorption dip ($\phi_{avg} \approx 0.80-0.845$) is present in the soft X-ray light curve of AR~UMa. The photoelectric nature of the dip is evidenced by its energy dependency, since $\tau_{\rm X}\propto \sigma(E)N_{H}$ (where $\sigma(E) \sim E^{-3}$ is the photoelectric cross-section over a wide range of X-ray energies E) which manifests in a highly pronounced absorption dip in soft X-rays that is completely absent in the hard band, Fig.~\ref{fig:pn_energyresolved}. Absorption by the stream is inevitable for systems in which the primary accreting pole is northern, given they also satisfy $\beta<<i$. However, in AR~UMa the primary accreting pole is southern and there is little evidence for an active northern pole, yet we observe the stream dip. For an inclination of 70$^\circ$, in order for the stream to obscure the emission region in the lower hemisphere while also preventing an eclipse, the stagnation region should have a large enough cross-sectional area. Such a puffed stagnation region has previously been invoked to explain the stream dip seen in UZ For \citep{UZ_For}. However, the mechanism that can lead to such large stream cross-sections at these distances is unknown, but may be connected to the stream–magnetosphere interaction described in Sect.~\ref{subsection:threadingshock}, wherein the shock-heated matter may expand. Assuming the absorbing matter is located in the stagnation region ($R_{\mu}^{\rm st}$), and adopting the system inclination of 70$^\circ$ we estimate the height of the intervening matter above the orbital plane to be $h$ = $R_{\mu}^{\rm st}$ $\rm cot(\rm 70^\circ)$ = 8.2 $\times$ 10$^{9}$ cm.

Perhaps the most interesting feature of the stream dip is its phase-jitter and substructure indicating a highly inhomogeneous nature of the accretion flow. For a cold absorber of solar abundance we measure column densities that vary in the broad range of $(0.2-1.3)\times 10^{22}$ cm$^{-2}$ from cycle to cycle. An estimate can be made of the distance between the X-ray emitting area and the absorption zone of the stream by substituting the observed duration of the stream-dip intervals ($\Delta\phi\sim0.04)$ in Eq.(4) of \cite{1995Watson} and is listed in Tab.~\ref{table:stream_parameters}. Further, using the per-cycle $N_{\rm H,stream}$, estimated in Sect. \ref{subsubsection:stream-dip}, we calculate the number density and in turn the instantaneous mass transfer rate using Eqs.~(6) and (7), respectively, from \cite{1995Watson} and are listed in Tab.~\ref{table:stream_parameters}. The median mass transfer rate estimated thus is $(2 \pm 1)\times10^{14}$gs$^{-1}$. The above estimates are based on the assumption of a smooth, homogeneous flow and thus fall short by almost two orders of magnitude compared to what is observed. Furthermore, if the absorption is indeed caused by the puffed-up stagnation region we are only observing (upper) part of the stream and not the total matter along the stream length which would also explain the low mass transfer rate determined using the observed stream density.

\begin{table}
\caption{Instantaneous mass transfer rates obtained from stream-dip width and interpolated column density from Tab.~\ref{table:stream_nH}.}
\label{table:stream_parameters}
\centering
\footnotesize
\begin{tabular}{ccccc}
\hline\hline
\noalign{\smallskip}
Cycle & $\Delta\phi$ & $d_{9}$ & $n_{14}$ & $\dot{M}$ \\
      &   & ($10^{9}\ \rm cm$) & ($10^{14}\ \rm cm^{-3}$) & ($10^{14}\ \rm g\ s^{-1}$) \\
\noalign{\smallskip}
\hline
\noalign{\smallskip}
1 & $0.058$ & $5.45$ & $0.63$ & $3.46$ \\
2 & $0.040$ & $7.88$ & $0.23$ & $1.04$ \\
3 & $0.040$ & $7.90$ & $0.10$ & $0.45$ \\
4 & $0.038$ & $8.38$ & $0.42$ & $1.84$ \\
5 & $0.035$ & $9.15$ & $0.46$ & $1.92$ \\
6 & $0.040$ & $7.96$ & $0.60$ & $2.72$ \\
\noalign{\smallskip}
\hline
\end{tabular}
\tablefoot{
Stream absorption parameters are derived by assuming $M_1 = 1.125\ M_\odot$ \citep{ARUMa_TESS}, stream-radius $r_s = 10^9\ \rm cm$, for an inclination $i=70^\circ$. Here, $\Delta\phi$ is the dip duration (i.e., FWHM of the dip profile), $d_{9}$ is the distance between the X-ray emission area and the absorbing part of the stream in units of $10^{9}\ \rm cm$, $n_{14}$ is the stream number density in $10^{14}\ \rm cm^{-3}$, and $\dot{M}$ is the estimated instantaneous mass-transfer rate.
}
\end{table}

\section{Conclusions} \label{conclusion}
AR~UMa, the highest field polar, shows an extreme case of soft blobby accretion with infinite soft excess. The mass transfer rate of the system is estimated to be $\dot{M} \approx (1.19 - 1.92) \times 10^{15} \ \mathrm{g\,s^{-1}}$, corresponding to $(1.18 - 3.05) \times 10^{-11} \ M_\odot\,\mathrm{yr}^{-1}$, within the median mass-transfer rate of polars, that is, $2.5\times10^{-11}\ M_\odot\ yr^{-1}$.
The accretion in the system occurs via dense, diamagnetic blobs or stream-aligned filaments that bury themselves within thick photospheric layers of the WD. The following qualitative signatures attest to this interpretation:
\begin{enumerate}
    \item The primary accreting pole exhibits exclusively soft X-ray emission, with no hard X-ray photons detected that would be attributable to a standing shock.
    \item The soft X-ray light curve shows flaring with dramatic cycle-to-cycle variability in its morphology.
    \item The accretion stream is porous, as evidenced by the variable density profile of the stream dip, which meanders in phase.
    \item A luminous threading shock with an average temperature of $\sim$1\,keV is detected, consistent with theoretical expectations.
\end{enumerate}

\begin{acknowledgements} 
 We thank the anonymous referee for their constructive comments that improved the clarity of the paper. This work is based on observations obtained with XMM-Newton, an ESA science mission with instruments and contributions directly funded by ESA Member States and NASA. KGP acknowledges support by the Deutsche Forschungsgemeinschaft under grant Schw536/37-2. We thank Jan Kurpas, Kira Knauff, and Laurent St\"utz, who obtained the photometric data at the 70-cm telescope.
\end{acknowledgements} 

\bibliographystyle{aa.bst}
\bibliography{references}
\newcounter{savedfigure}
\newcounter{savedtable}
\setcounter{savedfigure}{\value{figure}}
\setcounter{savedtable}{\value{table}}

\begin{appendix}

\FloatBarrier 
\clearpage
\onecolumn
\section{Blob characterization} \label{blob_appendix}
\begin{figure*}[h!]
    \centering
\includegraphics[width=0.7\textwidth]{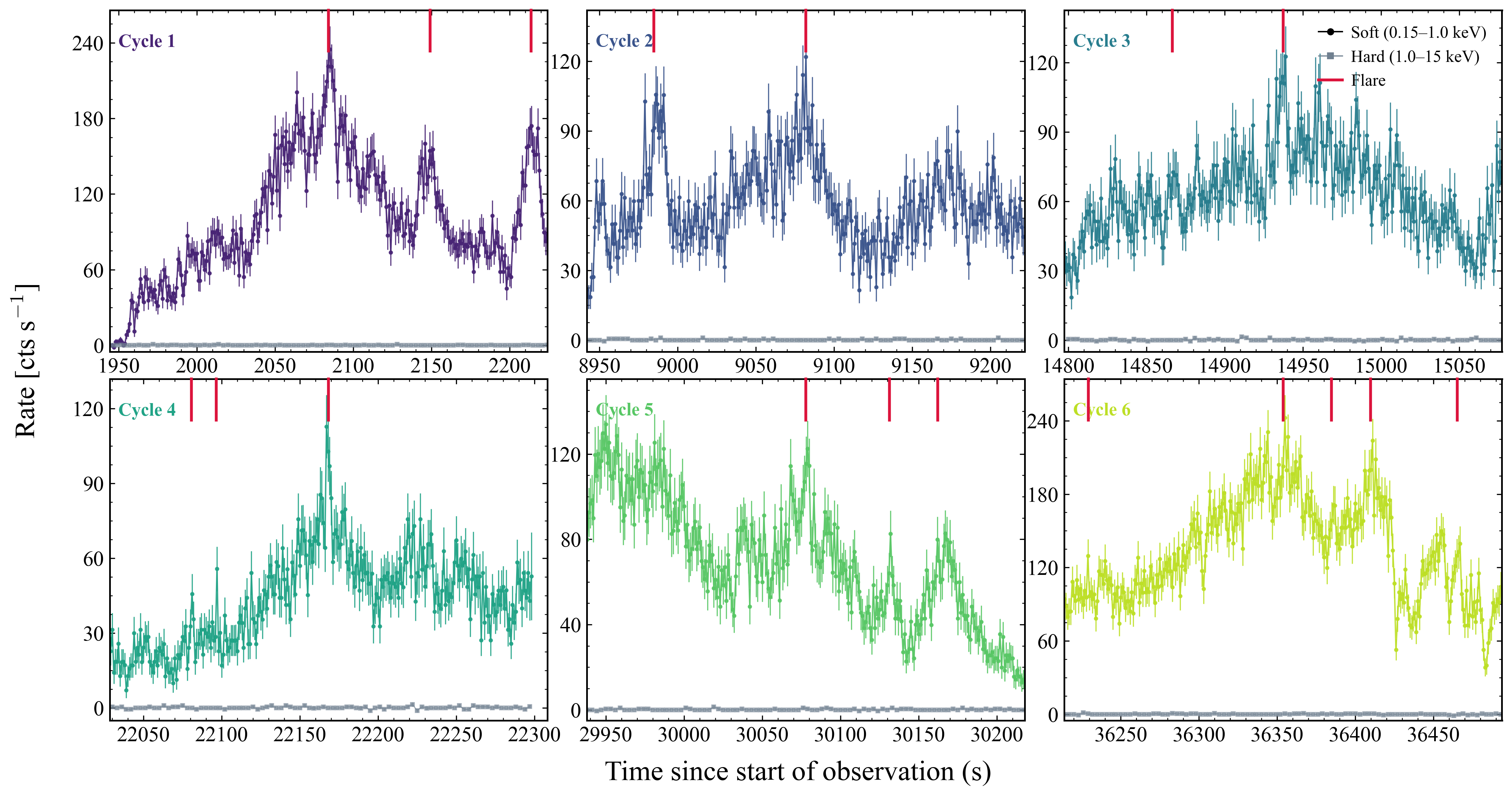}
        \caption{Some examples of resolved flares per cycle. The red line marks the identified flare. The colored lines show the EPIC pn soft X-ray (0.15-1 keV) light curve binned at 1~s resolution whereas the gray line show EPIC pn hard X-ray (1-15 keV) light curve binned at 3 s.}
    \label{fig:flare_cycle}
\end{figure*}
The blobs that originate during the stream-magnetospheric interaction discussed in Sect.~\ref{accretion_flow} are expected to have linear sizes of 0.01-0.1 $D_{\rm s}$, where $D_{\rm s}$ is the shock height estimated in Sect.~\ref{subsection:threadingshock}. For AR~UMa this corresponds to a filament length of $(7-70) \times 10^{6}$ cm. These filaments will get compressed as they move toward the WD by a factor of $(R_{\rm coup}/R_{\rm wd})^{1/2}$ which for AR UMa is approximately seven, implying a filament length at WD of $(5-50) \times 10^{7}$ cm. Therefore we expect the blackbody flux to produce a time variability of $\Delta t = \frac{l_{\rm wd}^{\rm blob}}{v_{\rm ff}}$, that is, $\approx$ 0.06-0.6 seconds. The predicted variability sets a Nyquist-required resolution of at least 0.3 seconds. However, at the observed source count rate estimated to be 27 counts s$^{-1}$ (for a 35 eV blackbody) a 0.3 second bin yields only eight counts (S/N of 2.8 $\sigma$) insufficient for statistically significant detection of a flare. We therefore choose a light curve of 1 second resolution (S/N 5$\sigma$) to investigate flaring events with the caveat that at this binning only the upper end of the variability will be measured and as a result the specific mass accretion rate determined will be lower as $\dot{m}\propto\frac{1}{\Delta t^{3}}$.

Within each cycle, discrete flares resolved against the undulating bright-phase level were identified, a few examples are shown in Fig. \ref{fig:flare_cycle}. The flare profiles are generally symmetric with an average rise ($\tau_{\rm rise}$) and decay time ($\tau_{\rm decay}$) of 10 seconds. For a distance of 98.62 pc and a geometric correction factor of 2$\pi$ we estimate the average energy released by a flare to be $\epsilon  = 3.2 \times 10^{33}$ erg, this corresponds to a blob mass $m_{\rm blob} = 9.6 \times 10^{15}$ g. Using the flare duration, $\Delta t$,  and the free-fall velocity, $v_{\rm ff}$, we estimate the average filament length at the WD surface $l_{\rm wd}^{\rm blob}$ to be $1.2 \times 10^{10}$ cm. In order to estimate the specific mass accretion rate and the pre-shock density of the blobs we need to have a realistic estimate of the instantaneous accreting area, $A_{\rm acc}$, which is physically distinct from, and in general vastly smaller than the effective area $A_{\rm eff}$ determining the X-ray spectral normalization \citep{HK_Area, 2000King}. However, as pointed out by \cite{2000King} no direct observational constraint on $A_{\rm acc}$ exists but it follows the strict hierarchy $A_{\rm acc} << A_{\rm eff} << 4\pi R_{\rm wd}^{2}$. Substituting $A_{\rm eff}$ for the unmeasurable $A_{\rm acc}$ therefore yields only a lower limit on the specific mass accretion rate, that is, $\dot m_{\rm LL}<<\dot m_{\rm true}$, since $A_{\rm acc}<<A_{\rm eff}$. This lower limit was evaluated per flare giving $\dot m_{\rm LL}$ in the range $\sim 0.7-4$\, g cm$^{-2}$ s$^{-1}$, far below the critical specific accretion rate $\dot m_{\rm crit}\sim10^{2}$\,g cm$^{-2}$ s$^{-1}$ required for the burial of shocks \citep{Kupiers_Pringle} and much lower than the estimates made in Sect.~\ref{accretion_flow}. In order to estimate a more realistic specific mass accretion rate we adopt the following heuristic approach. 
For a blob of length $l_{\rm coup}^{\rm blob}$ along the field and cross-sectional area $A_{\rm coup}^{\rm blob}$ at the coupling region ($R_{\rm coup}$), tidal stretching and magnetic compression along field lines converging onto the WD surface give $l_{\rm wd}^{\rm blob} = l_{\rm coup}^{\rm blob}\left(\frac{R_{\rm coup}}{R_{\rm wd}}\right)^{1/2}$ and $A_{\rm wd}^{\rm blob} = A_{\rm coup}^{\rm blob}\left(\frac{R_{\rm coup}}{R_{\rm wd}}\right)^{-3}$ \citep{1986HKL}. Assuming the clump is approximately isotropic at coupling ($l_{\rm coup}^{\rm blob}\sim d_{\rm coup}^{\rm blob}$, that is, $A_{\rm coup}\sim \frac{\pi}{4} ({l_{\rm coup}^{\rm blob}})^{2}$), the accreting cross-section at the WD surface purely in terms of the measured duration and the assumed coupling radius (here, $R_{\rm coup} \sim R^{st}_{\mu}$ \citep{Rcoup}), can be calculated as $A_{\rm wd}^{\rm blob} = \frac{\pi}{4}{l_{\rm wd}^{\rm \small{blob}}}^{\small 2} \left(\frac{R_{\rm wd}}{R_{\rm coup}}\right)^{4}$. The corresponding specific accretion rate for that blob is then $\dot m_{\rm blob} = \frac{m_{\rm blob}/\Delta t}{A_{\rm wd}^{\rm blob}} = \frac{4\ m_{\rm blob}}{\pi\ \Delta t^{3}\ v_{\rm ff}^{2}} \left(\frac{R_{\rm coup}}{R_{\rm wd}}\right)^{4}$. Unlike $\dot m_{\rm LL}$, $\dot m_{\rm blob}$ does not rely on $A_{\rm eff}$, however it instead carries its own modeling assumption, that is, the isotropic-clump approximation at coupling and the value adopted for $R_{\rm coup}$ (taken here to be the same as stream-magnetospheric radius). The specific mass accretion rate derived using $A_{\rm wd}^{\rm blob}$ estimates for the observed flares is in the range $(2.3-4000)$ g cm$^{-2}$ s$^{-1}$ with an average of 350 g cm$^{-2}$ s$^{-1}$, well within the shock burial threshold.

\begin{landscape}
\section{X-ray spectral analysis}
\subsection{EPIC best-fit parameters} \label{append:spectral_fits}
\begin{table}[h!]
\caption{X-ray spectral parameters for the faint and bright phases of AR~UMa.}
\label{tab:spectral_fits}
\centering
\resizebox{\linewidth}{!}{%
\renewcommand{\arraystretch}{1.3}
\begin{tabular}{llccccccccccc}
\hline\hline
Phase & Model & $N_{\rm H, tbabs}$ & $N_{\rm H, pcfabs}$ & $f_{\rm cov}$ & $kT_{\rm bb}$ & Norm$_{\rm bb}$ & $kT_{\rm apec1}$ & $kT_{\rm apec2}$ & $kT_{\rm apec3}$ & $\chi^2_{r}$ & $\log F_{\rm bb}$ & $\log F_{\rm plasma}$ \\
 & & ($10^{22}$ cm$^{-2}$) & ($10^{22}$ cm$^{-2}$) & & (eV)& & (keV) & (keV) & (keV) & & (cgs) & (cgs) \\
\hline
\textbf{EPIC MOS-1} \\
Faint & 2apec & $1.0 \times 10^{-3}$ & -- & -- & -- & -- & $1.1$ & $0.20$ & -- & $7.6$ & -- & -- \\
 & 3apec & $1.0 \times 10^{-3}$ & -- & -- & -- & -- & $1.0$ & $0.20$ & $10^*$ & $7.6$ & -- & -- \\
 & bbody+2apec & $1.0_{-1.0}^{+1.1} \times 10^{-3}$ & -- & -- & $28_{-2.9}^{+3.3}$ & $3.8_{-1.6}^{+2.9} \times 10^{-4}$ & $1.0_{-0.13}^{+0.12}$ & $0.26_{-0.26}^{+0.14}$ & -- & $0.69$ & $-10.49_{-0.05}^{+0.05}$ & $-12.11_{-0.05}^{+0.05}$ \\
 & bbody+3apec & $1.0_{-1.0}^{+1.1} \times 10^{-3}$ & -- & -- & $27_{-2.9}^{+3.4}$ & $3.9_{-1.4}^{+1.6} \times 10^{-4}$ & $0.95_{-0.05}^{+0.13}$ & $0.22_{-0.22}^{+0.15}$ & $10^*$ & $0.65$ & $-10.48_{-0.05}^{+0.05}$ & $-12.03_{-0.06}^{+0.05}$ \\
 & pcfabs*(bbody+3apec) & $1.2_{-1.2} \times 10^{-3}$ & $0.35_{-0.03}^{+0.04}$ & $0.95_{-0.26}^{+0.04}$ & $29_{-0.62}^{+0.58}$ & $7.2_{-0.76}^{+0.76} \times 10^{-3}$ & $0.93_{-0.09}^{+0.09}$ & $0.21_{-0.21}^{+0.03}$ & $10^*$ & $0.63$ & $-9.22_{-0.05}^{+0.04}$ & $-11.59_{-0.06}^{+0.05}$ \\
\hline
Bright & bbody+2apec & $1.0_{-1.0}^{+1.1} \times 10^{-3}$ & -- & -- & $34_{-0.56}^{+0.42}$ & $7.3_{-0.20}^{+0.85} \times 10^{-3}$ & $1.0^*$ & $0.26^*$ & -- & $1.5$ & $-9.21_{-0.01}^{+0.01}$ & $-12.01_{-0.07}^{+0.06}$ \\
& bbody+3apec & $1.0_{-1.0}^{+1.1} \times 10^{-3}$ & -- & -- & $34_{-0.50}^{+0.49}$ & $7.3_{-0.26}^{+0.80} \times 10^{-3}$ & $0.95^*$ & $0.22^*$ & $10^*$ & $1.4$ & $-9.21_{-0.01}^{+0.01}$ & $-11.93_{-0.07}^{+0.06}$ \\
 & pcfabs*(bbody+3apec) & $1.2 \times 10^{-3}$ & $0.35^*$ & $0.85_{-0.01}^{+0.01}$ & $33_{-0.31}^{+0.31}$ & $5.6_{-0.11}^{+4.0} \times 10^{-2}$ & $0.93^*$ & $0.21^*$ & $10^*$ & $0.90$ & $-8.33_{-0.01}^{+0.01}$ & $-11.56_{-0.07}^{+0.06}$ \\
 & bbody+3apec (0.25-10~keV) & $1.0_{-1.0}^{+1.0} \times 10^{-3}$ & -- & -- & $36_{-0.73}^{+0.87}$ & $5.3_{-0.26}^{+0.70} \times 10^{-3}$ & $0.95^*$ & $0.22^*$ & $10^*$ & $0.91$ & $-9.34_{-0.02}^{+0.01}$ & $-12.93_{-0.07}^{+0.06}$ \\
\hline  
\textbf{EPIC MOS-2} \\
Faint & bbody+3apec & $1.0 \times 10^{-3}$ & -- & -- & $23$ & $1.8\times 10^{-3}$ & $1.4$ & $0.7$ & $10^*$ & $2.8$ & -- & -- \\
& pcfabs*(bbody+3apec) & $1.2_{-1.2}^{+1.2} \times 10^{-3}$ & $1.9_{-0.8}^{+1.1}$ & $0.85_{-0.15}^{+0.1}$ & $29_{-2.1}^{+1.2}$ & $3.5_{-2}^{+5} \times 10^{-3}$ & $0.98_{-0.1}^{+0.1}$ & $0.4_{-0.3}^{+0.01}$ & $10^*$ & $1.3$ & $-9.5_{-0.04}^{+0.03}$ & $-11.4_{-0.04}^{+0.04}$ \\
& bbody+3apec (0.25-10~keV) & $1.0_{-1.0}^{+1.0} \times 10^{-3}$ & -- & -- & $27_{-0.4}^{+0.4}$ & $7.1_{-1.9}^{+1.9} \times 10^{-4}$ & $1.24_{-0.18}^{+0.3}$ & $0.6_{-0.3}^{+0.01}$ & $10^*$ & $1.3$ & $-10.2_{-0.14}^{+0.1}$ & $-12.1_{-0.04}^{+0.04}$ \\
\hline  
Bright & pcfabs*(bbody+3apec) & $1.2_{-1.2}^{+1.2} \times 10^{-3}$ & $1.9_{-0.8}^{+1.1}$ & $0.82_{-0.1}^{+0.1}$ & $36_{-1.0}^{+1.0}$ & $4.7_{-2.3}^{+4.6} \times 10^{-2}$ & $0.98^*$ & $0.4^*$ & $10^*$ & $2.0$ & $-8.3_{-0.08}^{+0.07}$ & $-11.4_{-0.05}^{+0.05}$  \\
 & bbody+3apec (0.25-10~keV) & $1.0_{-1.0}^{+1.0} \times 10^{-3}$ & -- & -- & $36_{-0.8}^{+0.8}$ & $7.1_{-0.8}^{+0.8} \times 10^{-3}$ & $1.24^*$ & $0.6^*$ & $10^*$ & $1.0$ & $-9.22_{-0.01}^{+0.01}$ & $-12.1_{-0.06}^{+0.05}$\\
\hline  
\textbf{EPIC pn} \\
  Faint & bbody+3apec & $1.0_{-1.0}^{+5.3} \times 10^{-3}$ & -- & -- & $25_{-1.3}^{+2.2}$ & $2.5_{-1.1}^{+2.7} \times 10^{-3}$ & $1.0_{-0.09}^{+0.09}$ & $0.27_{-0.05}^{+0.06}$ & $10^*$ & $0.97$ & $-9.77_{-0.05}^{+0.04}$ & $-11.93_{-0.03}^{+0.02}$ \\ 
    & pcfabs*(bbody+3apec) & $1.0_{-1.0}^{+1.0} \times 10^{-3}$ & $0.14_{-0.01}^{+0.01}$ & $0.97_{-0.05}^{+0.01}$ & $20_{-20}^{+0.15}$ & $0.42_{-0.04}^{+0.04}$ & $0.97_{-0.06}^{+0.06}$ & $0.23_{-0.02}^{+0.02}$ & $10^*$ & $0.94$ & $-7.59_{-0.04}^{+0.04}$ & $-11.77_{-0.03}^{+0.03}$ \\
   \hline
   Bright & bbody+3apec & $1.0 \times 10^{-3}$ & -- & -- & $34_{-0.90}^{+0.42}$ & $9.7_{-0.31}^{+3.0} \times 10^{-3}$ & $1.0^*$ & $0.27^*$ & $10^*$ & $1.2$ & $-9.13_{-0.01}^{+0.01}$ & $-11.91_{-0.04}^{+0.04}$ \\ 
   & pcfabs*(bbody+3apec) & $1.0 \times 10^{-3}$ & $0.14^*$ & $0.77_{-0.01}^{+0.17}$ & $30_{-0.17}^{+0.18}$ & $7.5_{-0.1}^{+0.6} \times 10^{-2}$ & $0.97^*$ & $0.23^*$ & $10^*$ & $1.06$ & $-8.25_{-0.01}^{+0.01}$ & $-11.77_{-0.04}^{+0.04}$ \\
   
   \hline\hline
Phase & Model & $kT_{\rm bb}$ & Norm$_{\rm bb}$ & $kT_{\rm apec1}$ & $kT_{\rm apec2}$ & $kT_{\rm apec3}$ & $E_{\rm line}$ & $\sigma_{\rm line}$ & Norm$_{\rm gauss}$ & $\chi^2_{r}$ & $\log F_{\rm bb}$ & $\log F_{\rm plasma}$ \\
 &  & (eV) & & (keV) & (keV) & (keV) & (keV) & (keV) & & & (cgs) & (cgs) \\
\hline
\textbf{RGS$\dagger$} \\

Faint & bbody+3apec+gauss & $28^*$ & $1_{-0.1}^{+0.1} \times 10^{-4}$ & $0.85_{-0.07}^{+0.07}$ & $0.16_{-0.01}^{+0.001}$ & $10^*$ & $- $ & $-$ & $-$ & $1.8$ & $-9.88^{+0.06}_{-0.07}$ &$-12.2^{+0.03}_{-0.04}$\\
Bright & bbody+3apec+gauss & $36.0_{-0.3}^{+0.3}$ & $1.36_{-0.04}^{+0.04} \times 10^{-3}$ & $0.85^*$ &  $0.158^*$ & $10^*$ & $0.404^*$ & $1.6_{-0.5}^{+0.9} \times 10^{-3}$ & $6.4_{-2.1}^{+2.1}\times 10^{-5}$ & 1.3 &$-9.94^{+0.01}_{-0.01}$& $-12.4^{+0.04}_{-0.03}$\\
\hline  
   \end{tabular}%
   }
\tablefoot{
\tablefoottext{*}{Frozen parameter.} \\
{$\dagger$}{The Galactic column density was frozen at 1 $\times$ 10$^{19}$ cm$^{-2}$.} \\
The errors on the parameters were obtained with the criterion of $\chi_{min}^{2}+ 2.73$ for 90\% confidence.}
\end{table}
\end{landscape}

\clearpage 
\twocolumn 

\subsection{Examining cycle-to-cycle variation of blackbody temperature using RGS} \label{append:cycle_bb_var}
\begin{figure}[ht!] 
        \centering
        \includegraphics[width=0.8\columnwidth]{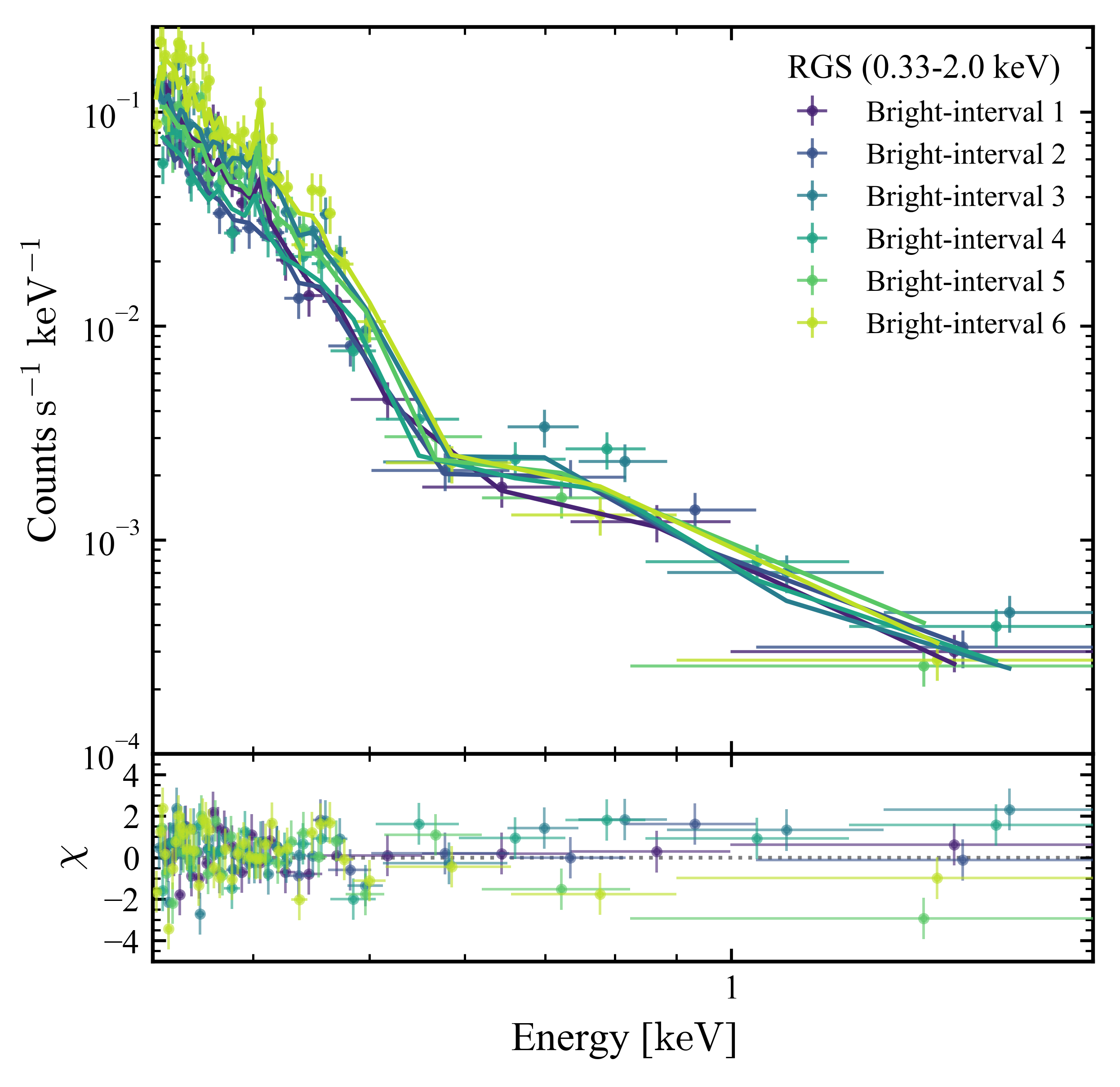}
         \caption{The RGS (0.33-2.0 keV), order-1 spectra and the best-fit model for the six bright intervals covered by the \textit{XMM-Newton} observation, bottom panel shows the residuals. }
    \label{fig:rgs-spec-6bp}
\end{figure}

We investigated possible cycle-to-cycle variation in the blackbody temperature of the accreting pole using high-resolution spectra from RGS (we have limited our analysis to order 1). We extracted spectra of bright interval of each cycle and fitted them jointly. While each cycle showed slightly different temperature for the soft component, they all agree within the errors. The best-fit temperatures obtained for each cycle are shown in Tab.~\ref{table:bb_temp_var} and Fig.~\ref{fig:rgs-spec-6bp} shows the RGS spectra with best-fit model and residuals.
\begin{table}[ht!]
\caption{Cycle-to-cycle variation in blackbody temperature ($kT_{\rm bb}$).}
\label{table:bb_temp_var}
\centering
\small
\begin{tabular}{cccc}
\hline\hline
\noalign{\smallskip}
Cycle & $kT_{\rm bb}$ & Norm$_{\rm bb}$ & $\chi^2_{r}$ \\
      & (eV) & ($10^{-4}$) & \\
\hline
1 & $34.2_{-2.1}^{+2.2}$ & $3.1_{-1.1}^{+1.9}$ & $0.94$ \\
2 & $36.3_{-2.4}^{+3.0}$ & $1.54_{-0.63}^{+0.97}$ & $0.66$ \\
3 & $38.3_{-1.7}^{+2.1}$ & $1.88_{-0.53}^{+0.63}$ & $1.2$ \\
4 & $38.3_{-2.3}^{+3.2}$ & $1.13_{-0.44}^{+0.54}$ & $1.2$ \\
5 & $38.0_{-2.0}^{+2.2}$ & $1.62_{-0.48}^{+0.67}$ & $1.2$ \\
6 & $35.9_{-1.5}^{+1.5}$ & $3.61_{-0.89}^{+1.30}$ & $1.65$ \\
\noalign{\smallskip}
\hline
\end{tabular}
\tablefoot{
The RGS order 1 (0.33-2.0 keV) was used to estimate the best fits, the model is defined as \texttt{Tbabs*(bbody+3apec+gauss)}. For all parameters except the blackbody temperature and normalization and Gaussian sigma and normalization the best-fit values obtained in Tab.~\ref{tab:spectral_fits} were kept frozen during the fit.
}
\end{table}
\subsection{Anomalous excess in RGS-1 spectrum} \label{rgs_anomaly_append}
\begin{figure}[ht!]
    \centering
    \includegraphics[width=0.8\columnwidth]{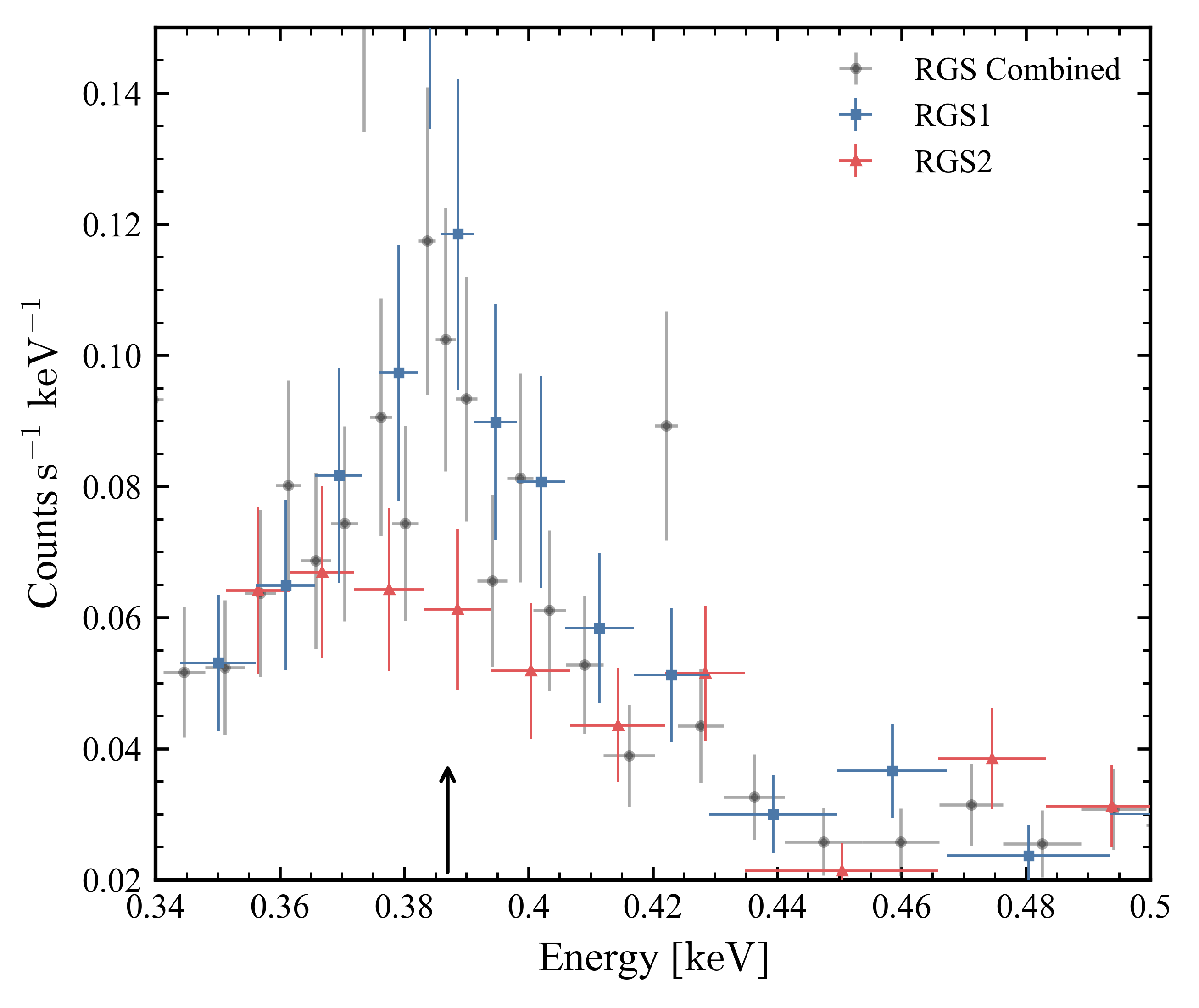}
    \caption{The RGS-1 spectrum showing anomalous excess at 0.387 keV skewing the combined RGS spectrum, the feature is missing in RGS-2. The cause of this is unknown but is potentially due to high dark current in CCD 2 in RGS-1.}
    \label{fig:rgs_anomaly}
\end{figure}

The RGS faint-phase spectrum reveals an anomalous excess centered around 0.387 keV in RGS-1. This excess is absent in RGS-2 and in the EPIC data, Fig.~\ref{fig:rgs_anomaly}. The cause of this feature is unknown but is potentially due to high dark current in CCD 2 of RGS-1 \footnote{\href{https://xmm-tools.cosmos.esa.int/external/xmm_user_support/documentation/uhb/rgsbkgd.html}{RGS background}}.

\section{Spectral energy distribution} \label{append:SED}
\begin{figure}[ht!]
    \centering  \includegraphics[width=\columnwidth]{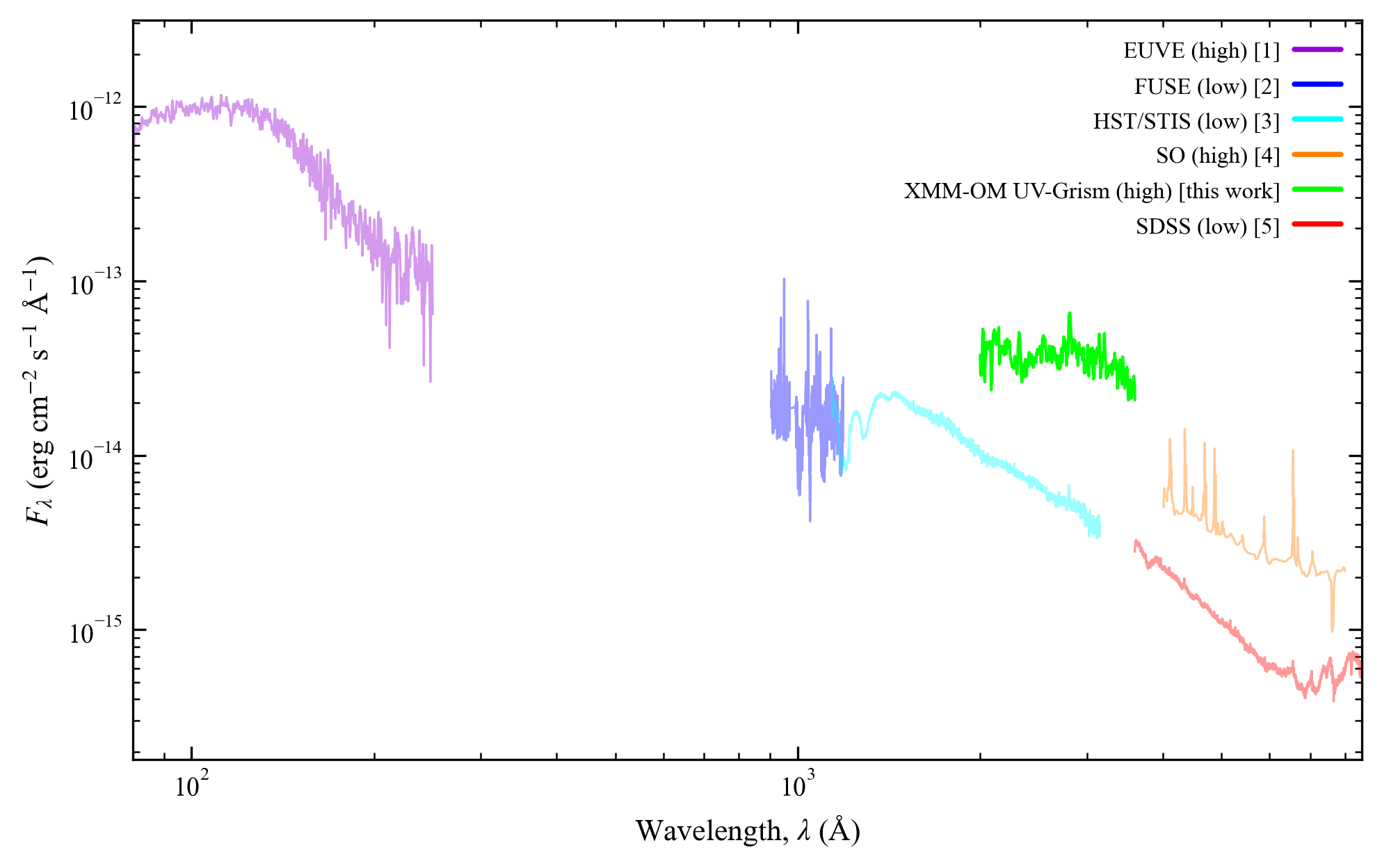}
    \caption{Spectral energy distribution (SED) of AR~UMa. The \textit{XMM-Newton} OM-UV Grism coadded spectrum is shown in green, the references for the remaining spectra are as follows; [1] \textit{EUVE}, high-state \cite{Szkody_EUVE_ASCA_RXTE_1999}, [2] \textit{FUSE}, low-state \cite{Hoard2004}, [3] \textit{HST} STIS, low-state \cite{STIS_ARUMa}, [4] \textit{SO:Steward Observatory}, high-state \cite{Schmidt_ARUMa_Geometry_1999}, [5] \textit{SDSS}-eBOSS spectrum obtained in 2012, low-state.}
    \label{fig:sed}
\end{figure}

Figure~\ref{fig:sed} shows the broad-band spectral energy distribution of AR~UMa. We combine archival low-state spectra; \textit{FUSE} (905-1187 $\AA$), \textit{HST} STIS (1150-3100 $\AA$), \textit{SDSS}-eBOSS (3600-10000 $\AA$), with high-state spectra; \textit{EUVE} (80-250 $\AA$), \textit{Steward Observatory (SO)} (4000-8000 $\AA$), and \textit{XMM-Newton} OM-UV Grism (2000-3600 $\AA$). The high-state UV continuum encompassed by the \textit{XMM-Newton} OM-UV Grism spectrum ($F_{\lambda}=(3.9\pm 0.8) \times 10^{-14}$ erg cm$^{-2}$ s$^{-1}$ \AA$^{-1}$), lies systematically above the archival low-state \textit{HST} STIS spectrum ($F_{\lambda} = (6.7 \pm 0.2) \times 10^{-15}$ erg cm$^{-2}$ s$^{-1}$ $\AA^{-1}$) across their overlapping wavelength range. The low spectral resolution of the OM UV Grism precludes resolving individual radiation components, but its continuum slope is markedly shallower than that of the \textit{HST} STIS spectrum, consistent with additional high-state UV contributions from the accretion stream and the irradiated secondary.

\end{appendix}

\end{document}